# Observation of Coherent Elastic Neutrino-Nucleus Scattering

**Authors:** D. Akimov[1,2], J.B. Albert[3], P. An[4], C. Awe[4,5], P.S. Barbeau[4,5], B. Becker[6], V. Belov[1,2], A. Brown[4,7], A. Bolozdynya[2], B. Cabrera-Palmer[8], M. Cervantes[5], J.I. Collar[9]*, R.J. Cooper[10], R.L. Cooper[11,12], C. Cuesta,[13], D.J. Dean[14], J.A. Detwiler[13], A. Eberhardt[13], Y. Efremenko[6,14], S.R. Elliott[12], E.M. Erkela[13], L. Fabris[14], M. Febbraro[14], N. E. Fields[9], W. Fox[3], Z. Fu[13], A. Galindo-Uribarri[14], M.P. Green[4,14,15], M. Hai[9], M.R. Heath[3], S. Hedges[4,5], D. Hornback[14], T.W. Hossbach[16], E.B. Iverson[14], L.J. Kaufman[3], S. Ki[4,5], S.R. Klein[10], A. Khromov[2], A. Konovalov[1,2], M. Kremer[4], A. Kumpan[2], C. Leadbetter[4], L. Li[4,5], W. Lu[14], K. Mann[4,15], D.M. Markoff[4,7], K. Miller[4,5], H. Moreno[11], P.E. Mueller[14], J. Newby[14], J.L. Orrell[16], C.T. Overman[16], D.S. Parno[13], S. Penttila[14], G. Perumpilly[9], H. Ray[17], J. Raybern[5], D. Reyna[8], G.C. Rich[4,14], D. Rimal[17], D. Rudik[1,2], K. Scholberg[5], B.J. Scholz[9], G. Sinev[5], W.M. Snow[3], V. Sosnovtsev[2], A. Shakirov[2], S. Suchyta[10], B. Suh[4,5,14], R. Tayloe[3], R.T. Thornton[3], I. Tolstukhin[3], J. Vanderwerp[3], R.L. Varner[14], C.J. Virtue[18], Z. Wan[4], J. Yoo[19], C.-H. Yu[14], A. Zawada[4], J. Zettlemoyer[3], A.M. Zderic[13]

**(COHERENT collaboration)**

* Correspondence to: collar@uchicago.edu

**Abstract**: The coherent elastic scattering of neutrinos off nuclei has eluded detection for four decades, even though its predicted cross-section is the largest by far of all low-energy neutrino couplings. This mode of interaction provides new opportunities to study neutrino properties, and leads to a miniaturization of detector size, with potential technological applications. We observe this process at a 6.7-sigma confidence level, using a low-background, 14.6-kg CsI[Na] scintillator exposed to the neutrino emissions from the Spallation Neutron Source (SNS) at Oak Ridge National Laboratory. Characteristic signatures in energy and time, predicted by the Standard Model for this process, are observed in high signal-to-background conditions. Improved constraints on non-standard neutrino interactions with quarks are derived from this initial dataset.

The characteristic most often associated with neutrinos is a very small probability of interaction with other forms of matter, allowing them to traverse astronomical objects while undergoing no energy loss. As a result, large targets (tons to tens of kilotons) are used for their detection. The discovery of a weak neutral current in neutrino interactions (*1*) implied that neutrinos were capable of coupling to quarks through the exchange of neutral Z bosons. Soon thereafter it was suggested that this mechanism should also lead to coherent interactions between neutrinos and all nucleons present in an atomic nucleus (*2*). This possibility would exist only as long as the momentum exchanged remained significantly smaller than the inverse of the nuclear size (Fig. 1A), effectively restricting the process to neutrino energies below a few tens of MeV. The enhancement to the probability of interaction (scattering cross-section) would however be very large when compared to interactions with isolated nucleons, approximately scaling with the square of the number of neutrons in the nucleus (*2, 3*). For heavy nuclei and sufficiently intense neutrino sources, this can lead to a dramatic reduction in detector mass, down to a few kilograms.

Coherent elastic neutrino-nucleus scattering (CEνNS) has evaded experimental demonstration for forty-three years following its first theoretical description. This is somewhat surprising, in view of the magnitude of its expected cross-section relative to other tried-and-tested neutrino couplings (Fig. 1B), and of the availability of suitable neutrino sources: solar, atmospheric and terrestrial, supernova bursts, nuclear reactors, spallation facilities, and certain radioisotopes (*3*). This delay stems from the difficulty in detecting the low-energy (few keV) nuclear recoil produced as the single outcome of the interaction. Compared to a minimum ionizing particle of the same energy, a recoiling nucleus has a diminished ability to generate measurable scintillation or ionization in common radiation detector materials. This is exacerbated by a trade-off between the enhancement to the CEvNS cross-section brought about by a large nuclear mass, and the smaller maximum recoil energy of a heavy target nucleus.

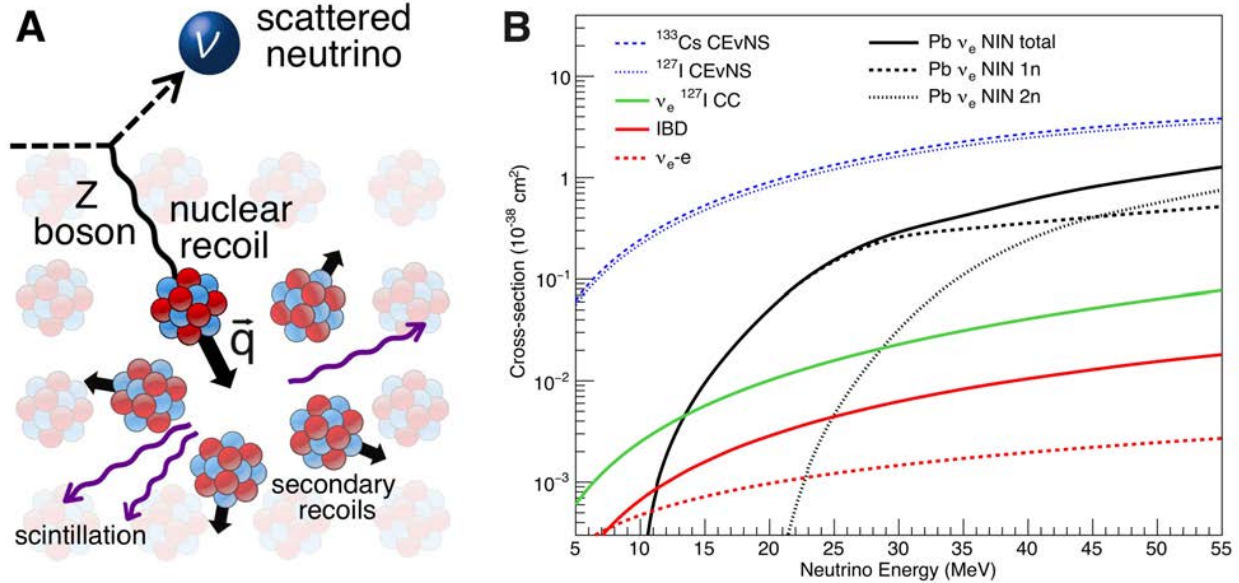

**Fig. 1**. (**A**) **Coherent Elastic Neutrino-Nucleus Scattering.** For a sufficiently small momentum exchange (q) during neutral-current neutrino scattering (qR < 1, where R is the nuclear radius in natural units), a long-wavelength Z boson can probe the entire nucleus, and interact with it as a whole. An inconspicuous low-energy nuclear recoil is the only observable. However, the probability of neutrino interaction increases dramatically, with the square of the number of neutrons in the target nucleus. In scintillating materials, the ensuing dense cascade of secondary recoils dissipates a fraction of its energy as detectable light. (**B**) **Total cross-sections from CEvNS and some known neutrino couplings.** Included are neutrino-electron scattering, charged-current (CC) interaction with iodine, and inverse beta decay (IBD). Because of their similar nuclear masses, cesium and iodine respond to CEvNS almost identically. The present CEvNS measurement involves neutrino energies in the range ~16-53 MeV, the lower bound defined by the lowest nuclear recoil energy measured (Fig. S9), the upper bound by SNS neutrino emissions (Fig. S2). The cross-section for neutrino-induced neutron (NIN) generation following $^{208}$Pb($\nu_e$,e$^-$ xn) is also shown. This reaction, originating in lead shielding around the detectors, can generate a potential beam-related background affecting CEvNS searches. The cross-section for CEvNS is more than two orders of magnitude larger than for IBD, the mechanism employed for neutrino discovery (*35*).

The interest in CEvNS detection goes beyond completing the picture of neutrino couplings predicted by the Standard Model of particle interactions. In the time since its description, CEvNS has been suggested as a tool to expand our knowledge of neutrino properties. These studies include searches for sterile neutrinos (*4-6*), a neutrino magnetic moment (*7, 8*), non-standard interactions mediated by new particles (*9-11*), probes of nuclear structure (*12*), and improved constraints on the value of the weak nuclear charge (*13*). In addition to these, the reduction in neutrino detector mass may lead to a number of technological applications (*14*), such as non-intrusive nuclear reactor monitoring (*15*). CEvNS is also expected to dominate neutrino transport in neutron stars, and during stellar collapse (*16-18*). Direct searches for Weakly Interacting Massive Particles (WIMPs), presently favored dark matter candidates, rely on the same untested coherent enhancement to the WIMP-nucleus scattering cross-section, and will soon be limited by an irreducible CEvNS background from solar and atmospheric neutrinos (*19*). The importance of this process has generated a broad array of proposals for potential CEvNS detectors: superconducting devices (*3*), cryogenic detectors (*20-22*), modified semiconductors (*23-25*), noble liquids (*26-30*), and inorganic scintillators (*31*), among others.

The Spallation Neutron Source (SNS) at Oak Ridge National Laboratory generates the most intense pulsed neutron beams in the world, produced by the interactions of accelerator-driven high-energy (~1 GeV) protons striking a mercury target. These beams serve an array of neutron-scattering instruments, and a cross-disciplinary community of users. Spallation sources are known to simultaneously create a significant yield of neutrinos, generated when pions, themselves a byproduct of proton interactions in the target, decay at rest. The resulting low neutrino energies are favorable for CEvNS detection (*3, 32, 33*). Three neutrino flavors are produced (prompt muon neutrinos $\nu_\mu$, delayed electron neutrinos $\nu_e$, and delayed muon anti-neutrinos $\bar{\nu}_\mu$), each with characteristic energy and time distributions (Fig. S2), and all having a similar CEvNS cross-section for a given energy. During beam operation, approximately $5 \times 10^{20}$ protons-on-target (POT) are delivered per day, each proton returning ~0.08 isotropically-emitted neutrinos per flavor. An attractive feature is the pulsed nature of the emission: 60 Hz of ~1 μs-wide POT spills. This allows us to isolate the steady-state environmental backgrounds affecting a CEvNS detector from the neutrino-induced signals, which should occur within ~10 μs windows following POT triggers. Similar time windows preceding the triggers can be inspected to obtain information about the nature and rate of steady-state backgrounds, which can then be subtracted (*31, 34*). A facility-wide 60 Hz trigger signal is provided by the SNS, at all times.

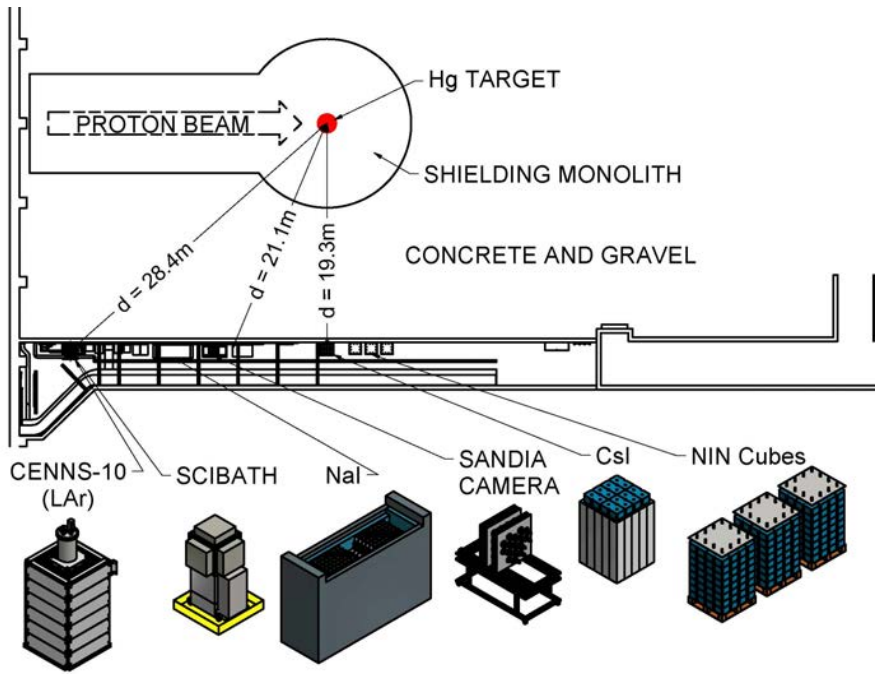

**Fig. 2**. COHERENT detectors populating the "neutrino alley" at the SNS (*34*). Locations in this basement corridor profit from more than 19 m of continuous shielding against beam-related neutrons, and a modest 8 m.w.e. overburden able to reduce cosmic-ray induced backgrounds, while sustaining an instantaneous neutrino flux as high as $1.7 \times 10^{11}$ $\nu_\mu$/cm$^2$ s.

As large as this neutrino yield may seem, prompt neutrons escaping the iron and steel shielding monolith surrounding the mercury target (Fig. 2) would swamp a CEvNS detector sited at the SNS instrument bay. Neutron-induced nuclear recoils would largely dominate over neutrino-induced recoils, making experimentation impossible. This led to a systematic investigation of prompt neutron fluxes within the SNS facility (*34*). A basement corridor, now dubbed the "neutrino alley" was found to offer locations with more than 12 m of additional void-free neutron-moderating materials (concrete, gravel) in the line-of-sight to the SNS target monolith. An overburden of 8 meters of water equivalent (m.w.e.) provides an additional reduction in backgrounds associated with cosmic rays. The CsI[Na] CEvNS detector and shielding described next were installed in the corridor location nearest to the SNS target (Fig. 2).

The advantages of sodium-doped CsI as a CEvNS detection material, its characterization for this application, and background studies using a 2 kg prototype are described in (*31*). Heavy cesium and iodine nuclei provide large cross-sections, and nearly-identical response to CEvNS (Fig. 1B), while generating sufficient scintillation for the detection of nuclear recoil energies down to a few keV. We performed supplementary calibrations of the final 14.6 kg CsI[Na] crystal before its installation at the SNS, as well as studies of the scintillation response to nuclear recoils in the relevant energy region (*34*). In addition to these, an initial dedicated experiment was performed at the chosen detector location, measuring the very small flux of prompt neutrons able to reach this position, and constraining the maximum contribution from the neutrino-induced neutron (NIN) background that can originate in lead shielding surrounding the detector (Fig. 1B, *34*). The conclusion from this measurement was that a CEvNS signal should largely dominate over beam-related backgrounds. The level of steady-state environmental backgrounds achieved in the final crystal slightly improved on expectations based on the prototype in (*31*), mostly thanks to refinements in data analysis, and to the presence of additional shielding. Further information about the experimental setup is provided in (*34*).

Figure 3 displays our main result, derived from fifteen months of accumulated live-time (Fig. S1). When comparing CsI[Na] signals occurring before POT triggers, and those taking place immediately after, we observe a high-significance excess in the second group of signals, visible in both the energy spectrum and the distribution of signal-arrival times. This excess appears only during times of neutrino production ("Beam ON" in the figure). The excess follows the expected CEvNS signature very closely, containing only a minimal contamination from beam-associated backgrounds (*34*). NINs have a negligible contribution, even smaller than that from prompt neutrons, which is shown in the figure. The formation of the excess is strongly correlated to the instantaneous power on target (Fig. S14). All neutrino flavors emitted by the SNS contribute to reconstructing the excess, as expected from a neutral current process. Stacked histograms in Fig. 3 display the Standard Model CEvNS predictions for prompt $\nu_\mu$ and delayed $\nu_e$, $\bar{\nu}_\mu$ emissions. Consistency with the Standard Model is observed at the one-sigma level (134 ± 22 events observed, 173 ± 48 predicted). A 2-D (energy, time) profile maximum likelihood fit favors the presence of CEvNS over its absence at the 6.7-sigma level (Fig. S13). Further details and a discussion of uncertainties are provided in (*34*), together with similar results from a parallel analysis (Fig. S11).

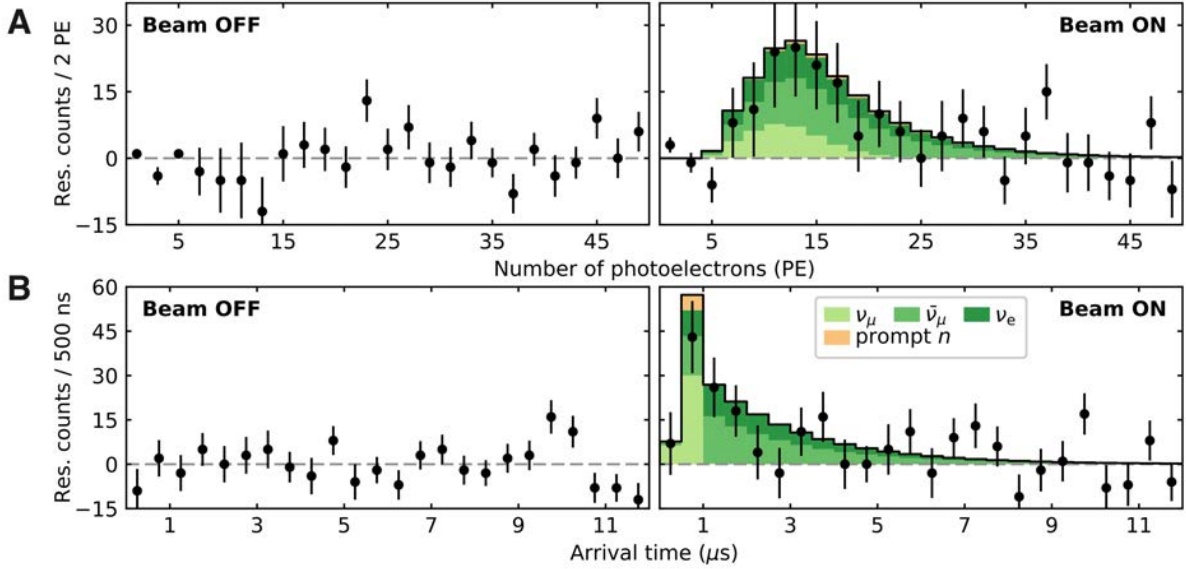

**Fig. 3. Observation of Coherent Elastic Neutrino-Nucleus Scattering.** Shown are residual differences (datapoints) between CsI[Na] signals in the 12 μs following POT triggers, and those in a 12-μs window before, as a function of their **(A)** energy (number of photoelectrons detected), and of **(B)** event arrival time (onset of scintillation). Steady-state environmental backgrounds contribute to both groups of signals equally, vanishing in the subtraction. Error bars are statistical. These residuals are shown for 153.5 live-days of SNS inactivity ("Beam OFF") and 308.1 live-days of neutrino production ("Beam ON"), over which 7.48 GWhr of energy (~1.76 x $10^{23}$ protons) was delivered to the mercury target. Approximately 1.17 photoelectrons are expected per keV of cesium or iodine nuclear recoil energy (*34*). Characteristic excesses closely following the Standard Model CEvNS prediction (histograms) are observed for periods of neutrino production only, with a rate correlated to instantaneous beam power (Fig. S14).

Figure 4 shows an example of CEvNS applications: improved constraints on non-standard interactions between neutrinos and quarks, caused by new physics beyond the Standard Model (*9-11*). These are extracted from the maximum deviation from Standard Model CEvNS predictions allowed by the present dataset (*34*), using the parametrization in (*30, 33*).

Data-taking continues, with neutrino production expected to increase this summer by up to 30%, compared to the average delivered during this initial period. In addition to CsI[Na], the

COHERENT collaboration presently operates a 28 kg single-phase liquid argon (LAr) detector, 185 kg of NaI[Tl] crystals, and three modules dedicated to the study of NIN production in several targets (Fig. 2). Presently planned expansion includes a ~1 ton LAr detector with nuclear/electron recoil discrimination capability, an already-in-hand 2 ton NaI[Tl] array simultaneously sensitive to sodium CEvNS and charged-current interactions in iodine (Fig. 1B), and p-type point contact germanium detectors (*24*) with sub-keV energy threshold. We intend to pursue the new neutrino physics opportunities provided by CEvNS using this ensemble.

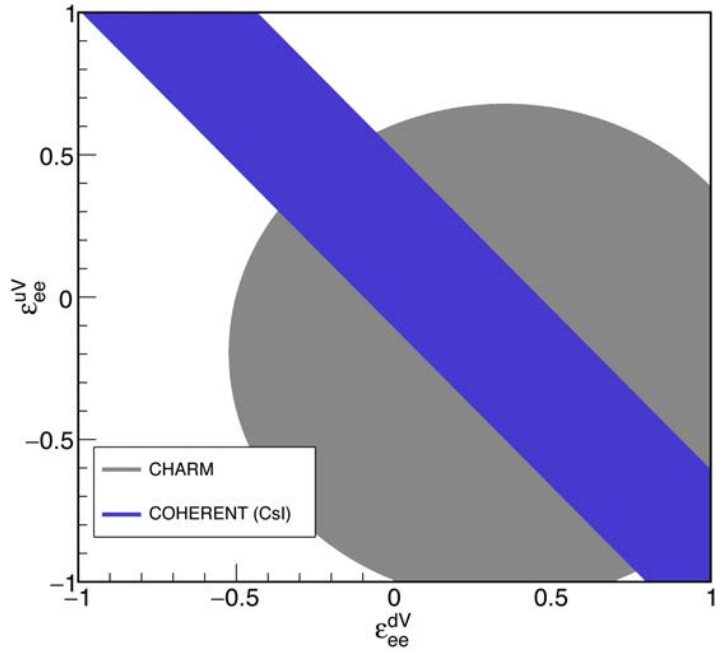

**Fig. 4**. **Constraints on non-standard neutrino-quark interactions**. Blue region: values allowed by the present data set at 90 % C.L. ($\chi^2_{min} < 4.6$) in $\varepsilon_{ee}^{uV}, \varepsilon_{ee}^{dV}$ space. These quantities parametrize a subset of possible non-standard interactions between neutrinos and quarks, where $\varepsilon_{ee}^{uV}, \varepsilon_{ee}^{dV} = 0,0$ corresponds to the Standard Model of weak interactions, and indices denote quark flavor and type of coupling. The gray region shows an existing constraint from the CHARM experiment (*34*).

**Acknowledgments:**

We acknowledge support from: the Alfred P. Sloan Foundation (BR2014-037), the Consortium for Nonproliferation Enabling Capabilities (DE-NA0002576), the Institute for Basic Science (Korea) (IBS-R017-G1-2017-a00), the National Science Foundation (PHY-1306942, PHY-1506357, PHY-1614545, HRD-1601174), Lawrence Berkeley National Laboratory Directed Research and Development funds, Russian Foundation for Basic Research (No. 17-02-01077_a), the Russian Science Foundation in the framework of MEPhI Academic Excellence Project (contract 02.a03.21.0005, 27.08.2013), Sandia National Laboratories Directed Research and Development Exploratory Express Funds, Triangle Universities Nuclear Laboratory, the U.S. Department of Energy Office of Science (DE-SC0009824, DE-SC0010007, Early Career Award DE-SC0014249, DE-SC0014558), the U.S. Department of Energy's National Nuclear Security Administration Office of Defense, Nuclear Nonproliferation Research, and Development, and the University of Washington Royalty Research Fund (FA124183).

Sandia National Laboratories is a multimission laboratory managed and operated by National Technology and Engineering Solutions of Sandia LLC, a wholly owned subsidiary of Honeywell International Inc. for the U.S. Department of Energy's National Nuclear Security Administration under contract DE-NA0003525.  Pacific Northwest National Laboratory is operated by Battelle for the United States Department of Energy under Contract No. DE-AC05-76RL01830, where work was supported through awards from the National Consortium for Measurement and Signature Intelligence Research Program, and from the Intelligence Community Postdoctoral Research Fellowship Program.

This work was supported in part by the Kavli Institute for Cosmological Physics at the University of Chicago through grant NSF PHY-1125897, and an endowment from the Kavli Foundation and its founder Fred Kavli.

This work was sponsored by the Laboratory Directed Research and Development Program of Oak Ridge National Laboratory, managed by UT-Battelle, LLC, for the U. S. Department of Energy. It used resources of the Spallation Neutron Source, which is a DOE Office of Science User Facility. This material is based upon work supported by the U.S. Department of Energy, Office of Science, Office of High Energy Physics.

This research used resources of the Oak Ridge Leadership Computing Facility, which is a DOE Office of Science User Facility.  Raw experimental data are archived at its High Performance Storage System, which provides 263 TB of COHERENT dedicated storage. We are grateful for additional resources provided by the research computing centers at the University of Chicago, and Duke University.

**Supplementary Materials**
Supplementary Text
Figures S1-S14
References *(36)-(84)*

# Supporting Online Material for

# Observation of Coherent Elastic Neutrino-Nucleus Scattering

D. Akimov[1,2], J.B. Albert[3], P. An[4], C. Awe[4,5], P.S. Barbeau[4,5], B. Becker[6], V. Belov[1,2], A. Brown[4,7], A. Bolozdynya[2], B. Cabrera-Palmer[8], M. Cervantes[5], J.I. Collar[9]*, R.J. Cooper[10], R.L. Cooper[11,12], C. Cuesta,[13]$, D.J. Dean[14], J.A. Detwiler[13], A. Eberhardt[13], Y. Efremenko[6,14], S.R. Elliott[12], E.M. Erkela[13], L. Fabris[14], M. Febbraro[14], N. E. Fields[9]¥, W. Fox[3], Z. Fu[13], A. Galindo-Uribarri[14], M.P. Green[4,14,15], M. Hai[9]¢, M.R. Heath[3], S. Hedges[4,5], D. Hornback[14], T.W. Hossbach[16], E.B. Iverson[14], L.J. Kaufman[3]†, S. Ki[4,5], S.R. Klein[10], A. Khromov[2], A. Konovalov[1,2]‡, M. Kremer[4], A. Kumpan[2], C. Leadbetter[4], L. Li[4,5], W. Lu[14], K. Mann[4,15], D.M. Markoff[4,7], K. Miller[4,5], H. Moreno[11], P.E. Mueller[14], J. Newby[14], J.L. Orrell[16], C.T. Overman[16], D.S. Parno[13]£, S. Penttila[14], G. Perumpilly[9], H. Ray[17], J. Raybern[5], D. Reyna[8], G.C. Rich[4,14]§, D. Rimal[17], D. Rudik[1,2], K. Scholberg[5], B.J. Scholz[9], G. Sinev[5], W.M. Snow[3], V. Sosnovtsev[2], A. Shakirov[2], S. Suchyta[10], B. Suh[4,5,14], R. Tayloe[3], R.T. Thornton[3], I. Tolstukhin[3], J. Vanderwerp[3], R.L. Varner[14], C.J. Virtue[18], Z. Wan[4], J. Yoo[19], C.-H. Yu[14], A. Zawada[4], J. Zettlemoyer[3], A.M. Zderic[13]

**COHERENT collaboration**

[1]Institute for Theoretical and Experimental Physics named by A.I. Alikhanov of National Research Centre "Kurchatov Institute", Moscow, 117218, Russian Federation.

[2]National Research Nuclear University MEPhI (Moscow Engineering Physics Institute), Moscow, 115409, Russian Federation.

[3]Department of Physics, Indiana University, Bloomington, IN 47405, USA.

[4]Triangle Universities Nuclear Laboratory, Durham, NC 27708, USA.

[5]Department of Physics, Duke University, Durham, NC 27708, USA.

[6]Department of Physics and Astronomy, University of Tennessee, Knoxville, TN 37996, USA.

[7]Department of Mathematics and Physics, North Carolina Central University, Durham, NC 27707, USA.

[8]Sandia National Laboratories, Livermore, CA 94550, USA.

[9]Enrico Fermi Institute, Kavli Institute for Cosmological Physics, and Department of Physics, University of Chicago, Chicago, IL 60637, USA.

[10]Lawrence Berkeley National Laboratory, Berkeley, CA 94720, USA.

[11]Department of Physics, New Mexico State University, Las Cruces, NM 88003, USA.

[12]Los Alamos National Laboratory, Los Alamos, NM 87545, USA.

[13]Center for Experimental Nuclear Physics and Astrophysics, and Department of Physics, University of Washington, Seattle, WA 98195, USA.

[14]Oak Ridge National Laboratory, Oak Ridge, TN 37831, USA.

[15]Department of Physics, North Carolina State University, Raleigh, NC 27695, USA.

[16]Pacific Northwest National Laboratory, Richland, WA 99352, USA.

[17]Department of Physics, University of Florida, Gainesville, FL 32611, USA.

[18]Department of Physics, Laurentian University, Sudbury, Ontario P3E 2C6, Canada.

[19]Department of Physics at Korea Advanced Institute of Science and Technology (KAIST) and Center for Axion and Precision Physics Research (CAPP) at Institute for Basic Science (IBS), Daejeon, 34141, Republic of Korea.

$ Now at Centro de Investigaciones Energéticas, Medioambientales y Tecnológicas (CIEMAT), Madrid 28040, Spain.

¥ Now at United States Nuclear Regulatory Commission, Lisle, IL 60532.

¢ Now at SpaceX Rocket Development Facility, McGregor, TX 76657.

† Now at SLAC National Accelerator Laboratory, Menlo Park, CA 94205, USA.

‡ Also at Moscow Institute of Physics and Technology, Dolgoprudny, Moscow region, 141700, Russian Federation.

£ Now at Department of Physics, Carnegie Mellon University, Pittsburgh, PA 15213, USA.

§ Also at Department of Physics and Astronomy, University of North Carolina at Chapel Hill, Chapel Hill, NC 27599, USA.

* Correspondence to: collar@uchicago.edu

**Supplementary Text**

**Experimental Setup: CEvNS Detector.** A 14.57 kg replica of the 2 kg prototype crystal described in (*31,36*) was procured from the same manufacturer (*37*), taking identical precautions in its encapsulation using low-background materials. The single difference between the detectors is a larger crystal length (34 cm vs. 4.7 cm). The photomultiplier (PMT) was upgraded to a low-background super bialkali Hamamatsu R877-100, which provides an enhanced quantum efficiency (*38*), resulting in a higher photoelectron (PE) yield. The shielding design described in (*31,36*) was upgraded to include an innermost 7.5 cm of high-density polyethylene (HDPE), selected for low radioactive content. This layer reduces the contributions from NIN backgrounds originating in the lead shield by more than an order of magnitude (*31*). Surrounding the HDPE is a 5 cm inner layer of low-activity lead (~10 Bq $^{210}$Pb / kg), followed by 10 cm of standard contemporary lead. A 5 cm-thick high-efficiency muon veto is placed around the sides and top of this gamma shield. The assembly is completed by a 15 cm HDPE pedestal, and water tanks containing > 9 cm of additional neutron moderator on the sides and top. A NI-5153 digitizer samples amplified CsI[Na] PMT signals and summed output from the muon veto at 500 MS/s, using the onset of protons-on-target (POT) as a trigger. This facility-wide 60 Hz trigger is shared by all SNS instruments, and maintained during Beam OFF periods. Collected traces are 70 μs long, spanning the CsI[Na] energy range from single photo-electrons (SPE) to ~400 keV in ionization (i.e., electron-equivalent) energy. Energies above ~60 keV saturate the digitizer range, and are accessible through an analysis of the long decay component of the scintillation.

The POT trigger position is set 55 μs into the digitized trace length. This allows us to analyze 12 μs regions before and after POT triggers in exactly the same fashion, using a 3 μs window to integrate the CsI[Na] light yield (expressed in number of PE) following the onset of a scintillation signal. The 40 μs-long windows preceding each of these two 12 μs segments are used to impose a data cut removing events caused by afterglow from a previous energy deposition (*31*). In what follows, we refer to these 40 μs windows as "pretraces", to signals and spectra in the 12 μs following POT triggers as "coincident" (C), and those in the 12 μs before (starting 15 μs prior to the POT trigger onset) as "anti-coincident" (AC). Figure S1 contains a number of system stability checks.

**Experimental Setup: SNS Neutrino Source**. Neutrinos at the SNS are a result of the decay of pions and muons created in a mercury target by a pulsed proton beam. Impinging on a dense target material, these protons rapidly lose energy via ionization, and secondarily via pion production. The high stopping power of mercury produces a compact interaction region along the beam, of ~15 cm. Pions are produced in proton-mercury interactions, and are quickly stopped in the target within ~3 nsec, with only a small probability of decay-in-flight (DIF). The stopped $\pi^-$ are captured on target nuclei, while stopped $\pi^+$ experience decay-at-rest (DAR), with production of mono-energetic 30 MeV $\nu_\mu$, referred to as "prompt" neutrinos. The $\mu^+$ from pion decay travels about one tenth of a millimeter and decays at rest with production of $\bar{\nu}_\mu$ and $\nu_e$, having a well-defined spectrum. These are referred to as "delayed" neutrinos. Both prompt and delayed neutrinos are emitted isotropically. Protons are produced at the SNS in short ~700 ns bursts, with a repetition rate of 60 Hz.

The neutrino production rate, as well as energy and time profiles (Fig. S2), are calculated using the Geant4 software package (*40*). The simulation includes the full geometry of the target, neutron moderators, surrounding beryllium reflector, neutrino time-of-flight (TOF), and time-profile of the POT pulse. The QGSP_BERT physics list is selected, which incorporates the Bertini (*41*) intra-nuclear cascade model of hadronic interactions.

Over the collected dataset, the SNS linear accelerator was operated at three different proton energies of 939.5, 957, and 973 MeV. This causes ~2 % variations in neutrino yield per proton. For the recorded data, we find an average production rate of 0.08 DAR neutrinos of each flavor per proton. Depending on assumptions made on the proton beam transverse profile, the resulting leakage of protons from the mercury target into the beryllium reflector generates < 10% variations in overall neutrino fluxes.

The same Bertini model, when implemented into the LAHET (*42*) software package, has been used to calculate the neutrino flux for the LSND and KARMEN experiments (*43-45*). In a later publication (*46*), containing a first estimation of the neutrino production for a SNS-like target, the LAHET code was renormalized to match low-energy data from (*47, 48*). This results in a lower neutrino production than what is obtained from Geant3 and Geant4 simulations. However, the latest global parameterization of pion production for proton-nucleon and nucleon-nucleon reactions (*49*) provides larger production cross-sections than these renormalized LAHET calculations. In view of these discrepancies, we assign a 10 % uncertainty to our neutrino flux predictions using Geant4-QGSP_BERT.

**Beam-Related Background Studies.** Prior to CsI[Na] experimentation, the full shield described in (*31,36*) was installed at the planned detector location, with the addition of the external water tanks mentioned above. An innermost 2.5 cm layer of ultra-low background lead was removed in order to house two 1.5 liter liquid scintillator (EJ-301) cells inside this shield. The cells were surrounded by a total of 2.2 tons of lead, with no intermediate neutron moderator. EJ-301 is primarily used for applications where neutron-gamma discrimination is required. Relying on the POT trigger signal, standard pulse-shape discrimination (PSD) techniques (*50, 51*) were applied to select EJ-301 neutron-like events taking place in the 10 $\mu$s before and after POT (Fig. S3). The cells were monitored by ET9390 PMTs, indicated for PSD applications (*51*). An optimal neutron-gamma separation and neutron signal acceptance were achieved in the 30-300 keV ionization energy range, the lower bound imposed by PSD limitations (*50, 51*), the upper bound by PMT saturation. Data were acquired for a total of 171.7 Beam ON live-days, over which 3.35 GWhr of proton energy was delivered to the SNS target.

The purpose of this initial experiment was to measure or constrain the two sources of beam-related backgrounds introduced in the main text: prompt SNS neutrons able to penetrate

19.3 m of moderating materials, and NINs. A very small interaction rate from the first was observed, at ~0.7 events per day of SNS beam operation (Fig. S3). The EJ-301 energy deposition spectrum from prompt neutrons (Fig. S4, top) was used to find a best-fit to their overall flux and spectral hardness (Fig. S4, bottom), described by a power law. This spectral choice, and the range of powers allowed by the fit, was based on previous results using the Scibath (*30*) and Sandia Camera (*52*) neutron detectors, sited at several locations along the "neutrino alley" (Fig. 2) and SNS instrument bay. These detectors were able to measure a prompt neutron flux of order $1.5 \times 10^{-7}$ n / cm$^2$ s (1-100 MeV) in the vicinity of the CsI[Na] location, albeit with a large uncertainty. The fitting procedure relied on an MCNPX-PoliMi (*53*) simulation of EJ-301 response in the 30-300 keV range, to an incoming flux of neutrons from the direction to the SNS target, bathing the full shielding geometry. Post-processing of the MCNPX-PoliMi output included statistical fluctuations in the generation of scintillation light, as well as the known response of EJ-301 to proton and carbon recoils (*54*). For proton recoil energies below 100 keV, the modified Lindhard model in (*55*) was adopted.

The resulting best-fit prompt neutron flux and spectral shape were propagated through a second MCNPX-PoliMi geometry, representative of the CsI[Na] detector and shield described in the first section of these supplementary materials. Post-processing of this simulation's output included Poisson fluctuations in photoelectron generation, signal acceptance generated by choice of data cuts (see "Data Analysis" below), and CsI[Na] response to nuclear recoils (i.e., its quenching factor, discussed in "Detector Calibrations"). Uncertainties in the determination of the prompt neutron flux and spectrum, as well as those associated with the quenching factor, were propagated through the simulation. The net outcome of this exercise was a predicted prompt neutron background rate in the CsI[Na] detector of 0.92 ± 0.23 events per GWhr of SNS energy delivery (Fig. 3 and Fig. S11). This background is highly-concentrated in time of arrival, and a factor of 25 smaller than the expected CEvNS signal rate.

An unbinned fit (56) to the arrival times of EJ-301 neutron-like signals (Fig. S3) was used to constrain NIN backgrounds, expected to arise dominantly via delayed $\nu_e$ through the $^{208}$Pb($\nu_e$,e$^-$ xn) reaction (57,58). The model employed for this fit is composed of random arrivals from environmental neutrons, the time-profile of the prompt neutron component (Fig. S3), and NINs following the $\nu_e$ time-profile (Fig. S2). The number of NIN events found by this fit is converted into a corresponding NIN production rate in lead, by means of an MCNPX-PoliMi simulation of EJ-301 response to a homogeneous and isotropic NIN emission from the lead shield. The energy spectrum of emitted NINs adopted by the simulation corresponds to the highest stellar-collapse neutrino energies treated in (59) (T = 8 MeV, i.e., $<E_\nu>$ = 25.2 MeV). While this $<E_\nu>$ is slightly softer than for SNS $\nu_e$'s (Fig. S2), we notice the negligible change in NIN spectral hardness with neutrino energy described in (59). The straggling time in lead for NINs on their way to the detectors is a few ns, and therefore neglected in fits and simulations. Post-processing of the simulation output, and propagation of uncertainties was done as above.

The NIN production rate obtained (0.97 ± 0.33 neutrons / GWhr / kg of Pb) is a factor 1.7 smaller than a prediction based on (57,58). This is compatible with the factor of ~3 uncertainty affecting theoretical predictions of neutrino cross-sections for heavy nuclei, in this energy range (60, 61). This production rate is used to generate simulated NINs uniformly in the CsI[Na] lead shield, with emission considerations, post-processing, and uncertainty propagation as above. The conclusion of this study is that the NIN background contamination affecting the present search is negligible: 0.54 ± 0.18 events / GWhr, a factor of ~47 smaller than the predicted CEvNS signal rate. It is ignored in what follows. The addition of 7.5 cm of HDPE internally to the CsI[Na] lead shield was determinant to achieve this (*31*).

The observed prompt neutron arrival times (Fig. S3) and best-fit spectral hardness (Fig. S4, bottom) were found to be in good agreement with their equivalent predictions from a Geant4 simulation of neutron production at the SNS target, and ensuing transport to the location of the detectors. In order to yield sufficient statistics, this simulation required the use of advanced Monte Carlo techniques (biasing, Russian roulette, particle splitting), in addition to parallelization on a large cluster, consuming ~300,000 CPU hours. This is a result of the large moderator thickness (19.3 m) modelled. While it is reassuring to observe agreement with Geant4 simulations, our conclusions on the expected rate of beam-related backgrounds do not rely on them. Separately, preliminary data from scintillator cells within NIN cubes, sited in close proximity to the CsI[Na] detector (Fig. 2), point at very similar constraints on the local prompt neutron flux, and comparable sensitivity to the NIN production rate in lead. Specifically, NIN cubes detect ~14.6 prompt neutrons per liter of EJ-301 per GWhr, compared to the ~11.9 n / liter GWhr in the present discussion, with only small differences in detectable fraction of proton recoil energies between both systems.

Weaker bounds on the maximum contribution from the prompt neutron background can be extracted directly from CEvNS search data. As described in (*31,36*), the inelastic scattering $^{127}$I(n,n'$\gamma$) reaction provides a convenient monitoring tool for fast neutrons, through its dominant 57.6 keV gamma emission. Figure S5 displays this spectral region for CsI[Na] signals in the 200-1100 ns arrival time range, collected over all Beam ON periods. The prompt neutron component derived from the EJ-301 analysis described above predicts just 1.2 ± 0.2 counts for this gamma peak. The fitting procedure described in the caption of Fig. S5 finds a number of counts (3.9 ±

11.1) compatible with zero, with a 90% confidence level (C.L.) upper limit of 22.2 counts. This upper bound is limited by the background achieved and presently collected exposure. A $^{252}$Cf neutron calibration, also shown in the same figure, provides a test of the MCNPX-PoliMi neutron transport simulations we have employed. The spectral hardness of $^{252}$Cf fission neutron energies above 1 MeV is comparable to the best-fit prompt neutron spectrum in Fig. S4, validating this choice of radioisotope. The simulated number of counts under the 57.6 keV peak during the 2 hr exposure to $^{252}$Cf is 662, with a 10% uncertainty carried from the manufacturer's yield calibration. A fit to the data (Fig. S5) finds a compatible 589 ± 68 counts.

**Detector Calibrations.** Three additional CsI[Na] detector studies beyond those in (*31*) are briefly described here. The first two are a gamma calibration of light yield uniformity using $^{241}$Am, and a $^{133}$Ba measurement of low-energy signal characteristics. These were performed on the 14.57 kg crystal used for the CEvNS search, following final potting of its PMT. The third consists of two new independent measurements of the CsI[Na] quenching factor. Those required the production of single nuclear recoils within the detector, and therefore involved a smaller 22 cm$^3$ crystal procured from the same manufacturer of the CEvNS detector (*37*), using an identical growth method and sodium dopant concentration (0.114 mole %).

**$^{241}$Am:** the light yield and light collection uniformity was measured using the 59.54 keV gamma emission from this isotope. This low energy ensures that interactions occur in the vicinity of a source placed on the external surface of the detector. This allows the investigation of local variations in light collection efficiency. Nine equally-spaced locations along the length of the scintillator were investigated. The average light yield for the eight closest positions to the PMT is

13.35 PE / keV, with individual fluctuations of ~0.5% (Fig. S6). The largest deviation was found at the most distant position, close to the back reflector, showing a negligible change by < 2 %.

$^{133}$**Ba:** the goal of this study was to produce a library of low-energy radiation-induced events with light yield below a few tens of PE, i.e., in the region of interest (ROI) for CEvNS (Figs. 3 and S11). To this end, a collimated pencil beam of $^{133}$Ba gamma rays was sent through a short path across the 14.57 kg crystal. Coincidences with a small backing gamma detector selected only forward-peaked Compton scattering events depositing a few keV of ionization energy in CsI[Na]. The resulting dataset contained negligible background contamination. The small differences in CsI[Na] scintillation decay time for nuclear and electronic recoils in this low-energy range (*31*), together with the use of a relatively short light-integration time (3 μs), enables us to employ $^{133}$Ba signals as close replicas of CEvNS events (Figs. S7 and S8). The library is employed to train data cuts aiming to discard certain spurious backgrounds: Cherenkov light emission in PMT window (*63*), random groupings of dark-current photoelectrons, misidentified scintillation onsets, etc. This is done while preserving a maximum of CEvNS-like radiation-induced signals (see "Data Analysis"). The data format and analysis procedure applied to $^{133}$Ba and CEvNS search data are identical: this allows use of the signal acceptance curve derived from this calibration (Fig. S9) for the generation of CEvNS signal and beam-related background predictions (see "CEvNS Signal Prediction and Statistical Analysis"). A first cut ("Risetimes" in Fig. S9) relies on the widely-used integrated rise-time method (*50, 51*), which consists of digitally constructing an integrated scintillation curve for each event, and finding the time difference between its crossing of two levels, defined as percent fractions of its maximum amplitude. It mainly removes a fraction of events with misidentified onsets, visible above the diagonals in Figs. S7 and S8. A second cut ("Cherenkov") demands a minimum number of

individual peaks in a scintillation signal, as determined by peak-finding algorithms. Its main effect is to remove random coincidences between Cherenkov light emission in the PMT window, and dark current or afterglow SPEs. Events resulting from this combination are often able to pass the "Risetimes" cut, generating the dominant steady-state background below ~20 PE (*63*).

**Quenching Factor (QF):** QF measurements establish the light yield from nuclear recoils, which are known to generate just a fraction of that produced by an electron recoil of similar energy. This fraction is expressed by the QF. We performed two new independent measurements using a monochromatic (3.8 ± 0.2 MeV) DD neutron beam (*64*) at the Triangle Universities Nuclear Laboratory (TUNL). Following interaction with the CsI[Na] target, scattered neutrons are detected at fixed angles, defining the nuclear recoil energy induced. The high purity and collimation of this beam allowed the use of a trigger condition based exclusively on scattered neutron detector signals. This helps avoid a number of known systematics able to bias QF measurements towards artificially large values at the smallest recoil energies. The crystal employed was instrumented with an ultra-bialkali PMT (*38*). This allowed the investigation of recoil energies down to 3 keV, lower than in any previous CsI[Na] QF measurement. The details of these calibrations are beyond the scope of this paper, and will be treated in a separate publication. Consistency with the method employed in the CEvNS search (3 µs light integration time, light yield non-linearity treatment as in (*65, 66*), using $^{241}$Am as a reference) was implemented in both beam calibrations, matching the energy scale used in the CEvNS search. An earlier measurement in (*31*) is excluded from the global fit of Fig. S10, as the different definition of QF used in that publication introduces a bias towards monotonically decreasing QF values with decreasing energy. This bias is caused by opposite trends in the dependence on energy of the scintillation decay constants from nuclear and electron recoils (*31*).

**Data Analysis.** Fig. S9 shows the effect of two additional data cuts, beyond those introduced in the $^{133}$Ba discussion. An "Afterglow" cut rejects signals having a certain number of peaks (typically SPEs, ≥ 4 in the figure) in their pretrace (see "Experimental Setup: CEvNS Detector"). Due to the very different levels of radiation sustained by the CsI[Na] crystal during $^{133}$Ba irradiation and CEvNS search, the signal acceptance for this cut is derived from the second. Sufficient removal of this background is of importance, as it has a modest effect in residuals like those in Figs. 3 and S11, but opposite to the formation of the CEvNS excess. "Quality" cuts remove three types of events specific to SNS data: muon veto coincidences, dead time from PMT saturation blocking by a linear gate, and digitizer range overflow. The stability of Quality and Afterglow cuts during SNS data-taking can be observed in Fig S1. The magnitude of event removal by the Afterglow cut, ~25% of the total, is necessary for an optimal signal-to-background ratio. This underlines the difficulty in performing this search using thallium-doped cesium iodide, a more common scintillator exhibiting much higher phosphorescence (*31*).

Two analysis pipelines were implemented. These treat SNS and calibration data independently already at the digitized trace level, applying separate reconstruction software and algorithms to extract analysis parameters: onset of signals, integrated charge, number of peaks, rise-times, gain stability corrections, etc. The first[1] (Fig. 3) used a comparison between Beam OFF AC data, by definition containing steady-state environmental backgrounds only, and CEvNS signal rate predictions, with the goal of determining the choice of cuts that maximizes the signal-to-background ratio. The rest of SNS data (Beam OFF C, Beam ON AC and C), were analyzed only once this choice was frozen, implementing a form of blind analysis. The second[2] employed a different approach to cut optimization, exercised on Beam ON AC data only, and independent of CEvNS predictions. It relied on maximization of the ratio of event acceptance in an energy ROI (4-20 PE), to the number of background events passing the same cuts. Both point at nearly identical best choices for dominant Cherenkov and Afterglow cuts, resulting in comparable residuals that contain CEvNS-like excesses for Beam ON periods only (Fig. 3 and Fig. S11). For sub-optimal cut choices in both pipelines, we observe continued good agreement between Beam ON C-AC residuals and CEvNS predictions, even if, as expected, the statistical evidence for CEvNS deteriorates.

---

[1] Centralized at the University of Chicago.

**CEvNS Signal Prediction and Statistical Analysis.** A two-dimensional energy and time ($6 \leq$ PE $< 30$, $< 6$ μs), binned, maximum likelihood estimator was used to fit Beam ON coincident (C) data (Fig. S12) to probability distribution functions (PDFs) for the CEvNS signal, the prompt neutron background, and steady-state environmental backgrounds. NIN backgrounds were omitted due to their small contribution (4 counts, spread out over ~6 μs and ~25 PE). The formulation of the prompt neutron PDF, its expected rate, and uncertainties, are implemented following their discussion in "Beam-Related Background Studies". The steady-state background PDF was generated from Beam ON AC data (Fig. S12): treating the background dimensions (energy, time) as uncorrelated, an analytical model was fit to its time distribution, following integration over energy. This model describes the shape of this PDF as a function of event arrival time. The corresponding time-integrated distribution describes the shape of the PDF in the energy dimension. The resulting two-dimensional shape was used as the background model.

---

[2] Centralized at the University of Tennessee and ITEP/MEPhI, Moscow.

The CEvNS PDF was generated by convolving the simulated neutrino flux (Fig. S2) at the CsI[Na] detector position with the CEvNS differential cross-section described in (*70*), including axial and vector couplings, as well as the strange quark contributions described in (*71*). Small (few %) differences in the CEvNS cross-section across flavors, arising from the neutrino charge radius (*72*), are neglected. Nuclear form factors (for Cs and I) are as described in (*73*). The PDF also accounts for the light yield of the crystal (Fig. S6), the quenching factor (Fig. S10), as well as Poisson fluctuations expected to affect the number of observed photoelectrons. The optimized signal acceptance curve (Fig. S9) is applied to both prompt neutron and CEvNS PDFs. The top panels in Fig. S13 display the 2-D surfaces of the PDFs employed by the fit.

The effect on the CEvNS PDF spectral shape arising from uncertainties in signal acceptance (Fig. S9), choice of form factor (*26*, *73*, *74*), and quenching factor, is found to be negligible. Uncertainties due to light yield uniformity (Fig. S6), and source-detector distance - measured using surveying techniques- are also negligible. The non-negligible uncertainty on the integrated CEvNS signal counts arises from: signal acceptance (generating a ± 5 % uncertainty), choice of form factor (± 5 %), neutrino flux (see "Experimental Setup: SNS Neutrino Source") (± 10 %), and quenching factor (± 25 %) (Fig. S10), for a total uncertainty of ± 28 %. The energy-independent quenching factor in Fig. S10 is conservative, as it gives rise to a larger uncertainty on the CEvNS signal rate than QF models following the data from the individual experiments shown there. The amplitude of the CEvNS signal is left unconstrained in the fit. The uncertainty on the prompt neutron rate (± 25 %) resulting from our knowledge of the prompt neutron flux, spectral hardness (Fig. S4), and quenching factor, was included as a constraint in the fit. The amplitude of the steady-state background PDF, and its constraint, were determined from the AC data.

The best-fit CEvNS signal obtained by the profile likelihood fit (*75*) is 134 ± 22 counts, which is 77 ± 16 per cent of the value predicted by the Standard Model (Fig. S13). The alternative hypothesis (presence of CEvNS) is favored over the null hypothesis at the 6.7 sigma level. Coverage tests were performed by means of a toy Monte Carlo, revealing no bias in the best-fit CEvNS signal counts.

A more simplistic analysis of the excess counts in the Beam ON C data (547 counts) over the AC data (405 counts), within the same two-dimensional window can also be performed, though it does not benefit from the knowledge of any energy spectra or arrival time information. It assigns a similar 136 ± 31 counts to the CEvNS signal.

The statistical analyses described above refer to data from the first analysis pipeline (Figs. 3 and Fig. S12). When applied to data from the second pipeline (Fig. S11), very similar conclusions are derived.

In addition to these tests, Fig. S14 shows the growth of the CEvNS-like excess observed, plotted next to cumulative SNS beam activity, normalized to the same vertical scale. There is a strong correlation between the excess in Beam ON C data and beam power, consistent with its being entirely due to beam-induced events. This was verified through repeated application of a Kolmogorov-Smirnov test (*76*), modified for background-subtracted data and calibrated using Monte Carlo-generated distributions of events. The SNS data showed a stronger correlation than 96% of the simulated distributions, indicating an absence of bursts or other time-varying anomalies. A similar test showed no anomalies in the Beam OFF data behavior.

As a final remark in this section, we call attention to the larger predicted CEvNS signal formation rate in (*31*). Its calculation followed a nearly-identical procedure to that above, but it employed an optimistic SNS neutrino production rate of 2.97 x $10^{22}$ ν / flavor / year, traceable to

(*77*). That reference assumes continuous SNS operation, a beam power not yet delivered (1.4 MW), and significantly higher energy per proton (1.3 GeV) and neutrino yield per proton (0.13) than actual values. The neutrino production rate during the first calendar year in this dataset was a factor ~3.75 smaller than what is derived from (*77*).

**Bounds on Non-Standard Interactions (NSI) Between Neutrinos and Quarks.** New physics that is specific to neutrino-nucleon interactions is currently quite poorly constrained, and is motivated in some beyond-the-Standard-Model scenarios (e.g., *9-11*). The existence of such non-standard interactions of neutrinos could confound neutrino mass ordering determination by long-baseline neutrino oscillation experiments (*79-81*). A constraint of these possibilities by scattering experiments is highly desirable in the context of the global neutrino physics program.

Here we employ the model-independent parameterization of non-standard contributions to the neutrino-quark interaction cross section, following (*33, 70, 81*), in which non-zero values of the $\varepsilon_{fg}^{qP}$ parameters describe either ``non-universal'' ($f = g$) or flavor-changing ($f \neq g$) interactions of neutrinos with quark flavor $q$, and $P = V$ (vector coupling), $A$ (axial coupling). The presence of non-zero NSI can enhance or suppress the CEvNS rate; if one neglects axial contributions and in the approximation $T \ll E_\nu$, where $T$ is the nuclear recoil energy and $E_\nu$ is the neutrino energy, the presence of non-zero NSI results in an overall scaling of the event rate, either enhancement or suppression, rather than a spectral distortion. A scattering experiment using neutrinos from pion decay at rest will have sensitivity only to NSI parameters with $f = e, \mu$ (i.e., all but $\varepsilon_{\tau\tau}^{qP}$). Here, as an example of a constraint analysis on parameters which are currently poorly known, we consider only non-zero values of $\varepsilon_{ee}^{uV}, \varepsilon_{ee}^{dV}$. We assume that the standard three-flavor model of neutrino mixing holds, and that the baseline is too short for significant flavor transition.

We treat the measurement as a single-bin counting experiment and perform a simple pull test (*82*). Given a measured number of steady-state-background-subtracted counts $N_{meas}$, which is 142 (547 beam ON minus 405 AC) for our current sample (see "CEvNS Signal Prediction and Statistical Analysis"), we can compare this to the predicted $N_{NSI}(\varepsilon_{ee}^{uV}, \varepsilon_{ee}^{dV})$ corresponding to a specific exposure and flux. We define a $\chi^2$ as follows:

$$\chi^2 = \frac{(N_{meas} - N_{NSI}(\varepsilon_{ee}^{uV}, \varepsilon_{ee}^{dV})[1+\alpha] - B_{on}[1+\beta])^2}{\sigma_{stat}^2} + \left(\frac{\alpha}{\sigma_\alpha}\right)^2 + \left(\frac{\beta}{\sigma_\beta}\right)^2,$$

where:

- $\sigma_{stat} = \sqrt{N_{meas} + 2B_{ss} + B_{on}}$ is the statistical uncertainty.
- $B_{on}$ is the estimated beam-on background. Including prompt neutrons but ignoring NINs, we estimate $B_{on} = 6$ for our current exposure, as described in "Beam-Related Background Studies".
- $B_{ss}$ is the estimated steady-state background (determined with AC data). We assume no systematic uncertainty on this estimate. For the current exposure, $B_{ss} = 405$, as in the simplistic analysis described in "CEvNS Signal Prediction and Statistical Analysis".
- $\alpha$ is the systematic parameter corresponding to uncertainty on the signal rate. $\sigma_\alpha$ is the fractional uncertainty corresponding to a 1-sigma variation. We estimate $\sigma_\alpha = 0.28$, incorporating flux, form factor, QF and signal acceptance uncertainties, as described in "CEvNS Signal Prediction and Statistical Analysis". In principle, flux, QF and acceptance uncertainties could affect $B_{on}$ as well, but in this case, as the Beam ON background has been estimated using data rather than the flux estimate, there is no flux-related uncertainty on $B_{on}$. QF and acceptance uncertainties on $B_{on}$ are neglected.

- $\beta$ is the systematic parameter corresponding to uncertainty on the estimate of $B_{on}$, uncorrelated with signal uncertainty. $\sigma_\beta$ is the fractional uncertainty corresponding to 1-sigma variation. We adopt $\sigma_\beta = 0.25$, as given in "Beam-related Background Studies"; it is assumed to be dominated by determination of the prompt neutron flux and spectrum.

This $\chi^2$ is minimized over the systematic nuisance parameters $\alpha, \beta$, for each value of $(\varepsilon_{ee}^{uV}, \varepsilon_{ee}^{dV})$. The results are shown in Fig. 4. The global best-fit values are $\varepsilon_{ee}^{uV} = -0.57, \varepsilon_{ee}^{dV} = 0.59, \alpha = 0.02, \beta = 0.02$. We note that the $\chi^2$ map is flat along the bands in $\varepsilon_{ee}^{uV}, \varepsilon_{ee}^{dV}$ -space corresponding to NSI event-rate suppression, so the specific best-fit values within the allowed region are sensitive to assumptions. We also note that according to reference (*80*), the constraints from the CHARM experiment (*81, 83*) may apply only for heavy new-interaction mediators, and that new interactions with light mediators are unconstrained by the CHARM data. Even with this first COHERENT data set, NSI parameters are meaningfully constrained.

We note that a measurement employing more than one target nucleus, as planned within the COHERENT collaboration, will enable more stringent constraints on the couplings; the more the $(A + N)/(A + Z)$ ratio differs between the targets, the better, as shown for example in (*33, 84*).

**Figures S1-S14**

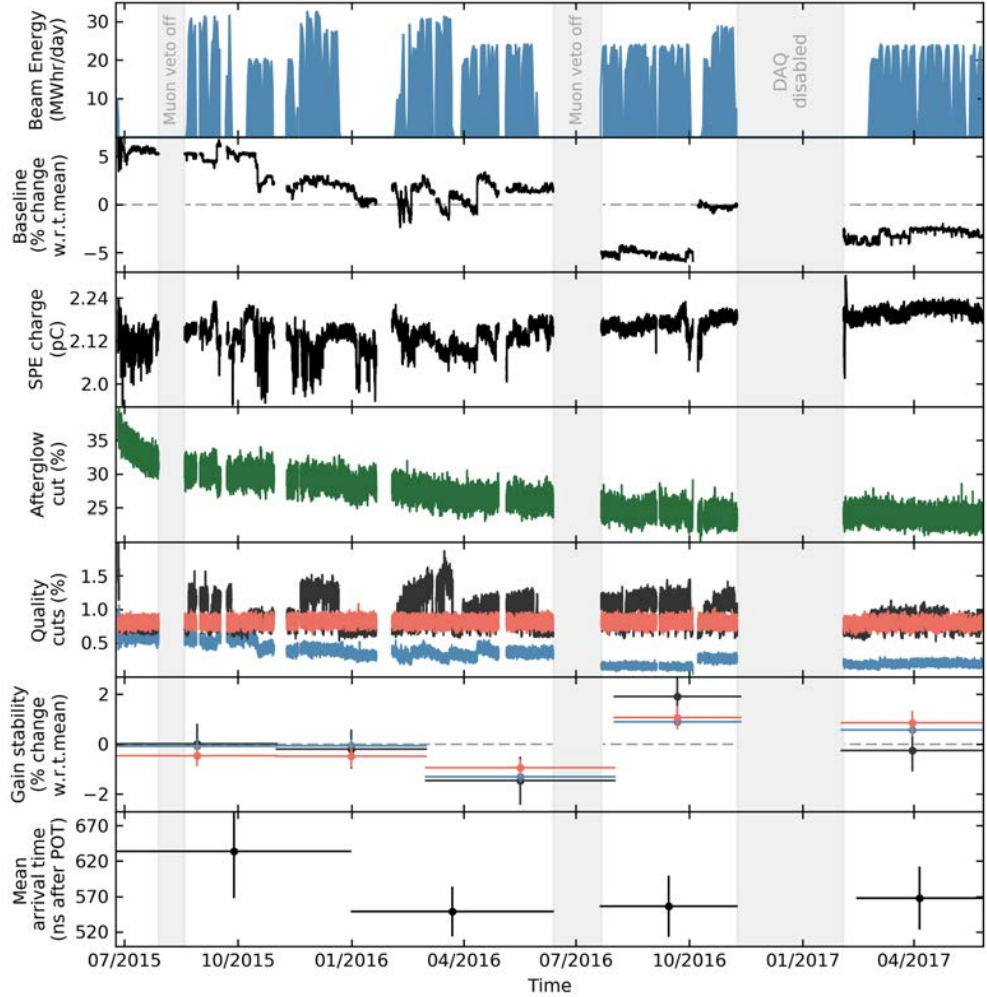

**Fig. S1**. Stability tests performed during the data-taking period. Top to bottom: 1) Daily energy delivered to the SNS target. Vertical grayed bands indicate periods of data loss. 2) Ten-minute average change in DC-level of the CsI[Na] channel baseline. 3) Ten-minute average of the integrated SPE charge. Event energies are normalized to this value, correcting for small PMT gain fluctuations. 4) Percentage of events removed by the Afterglow cut, proportional to steady-state background rate (see "Data Analysis" section). The observed decay is compatible with cosmogenic $^{125}$I production (*39*) and thermal neutron capture in $^{133}$Cs. 5) Percentage of events removed by Quality cuts (see "Data Analysis" section). Black: events rejected by coincidences with muon veto panels. The correlation with energy on target arises from the proximity to MOTS system exhaust gases (Fig. S12). Red: *idem* for linear gate cut. Blue: *idem* for data-acquisition range overflow. 6) Gain stability derived from the measured energy of $^{212}$Pb, $^{214}$Pb gamma backgrounds internal to the crystal (*31*) (black = 239 keV, red = 295 keV, blue = 352 keV). 7) Stability of the SNS POT trigger signal, extracted from prompt neutron interactions with muon veto panels. These cover a large 3 m$^2$ area, but have limited neutron efficiency due to their high trigger threshold. Neutron arrival times are similar to those in Fig. S3.

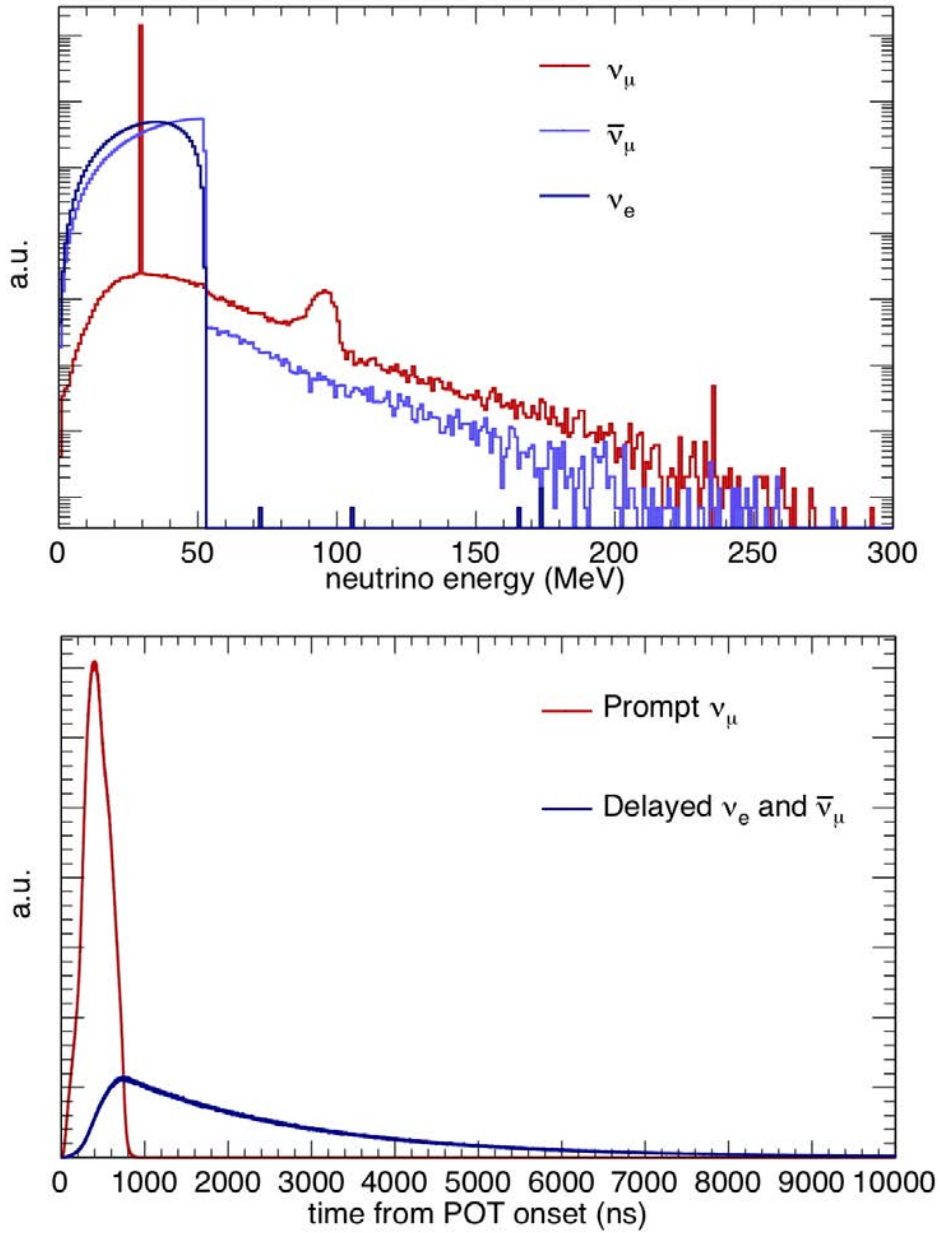

**Fig. S2**. Geant4 energy distribution and arrival time of SNS neutrinos to the CsI[Na] detector. Neutrinos above the endpoint of the Michel spectrum (~53 MeV) arise from DIF and muon capture, contributing a negligible (< 1%) signal rate. Delayed neutrinos follow the 2.2 μs time constant characteristic of muon decay. A discussion on neutrino production rates (the normalization factors for these distributions) and associated uncertainties is provided in the supplementary materials text.

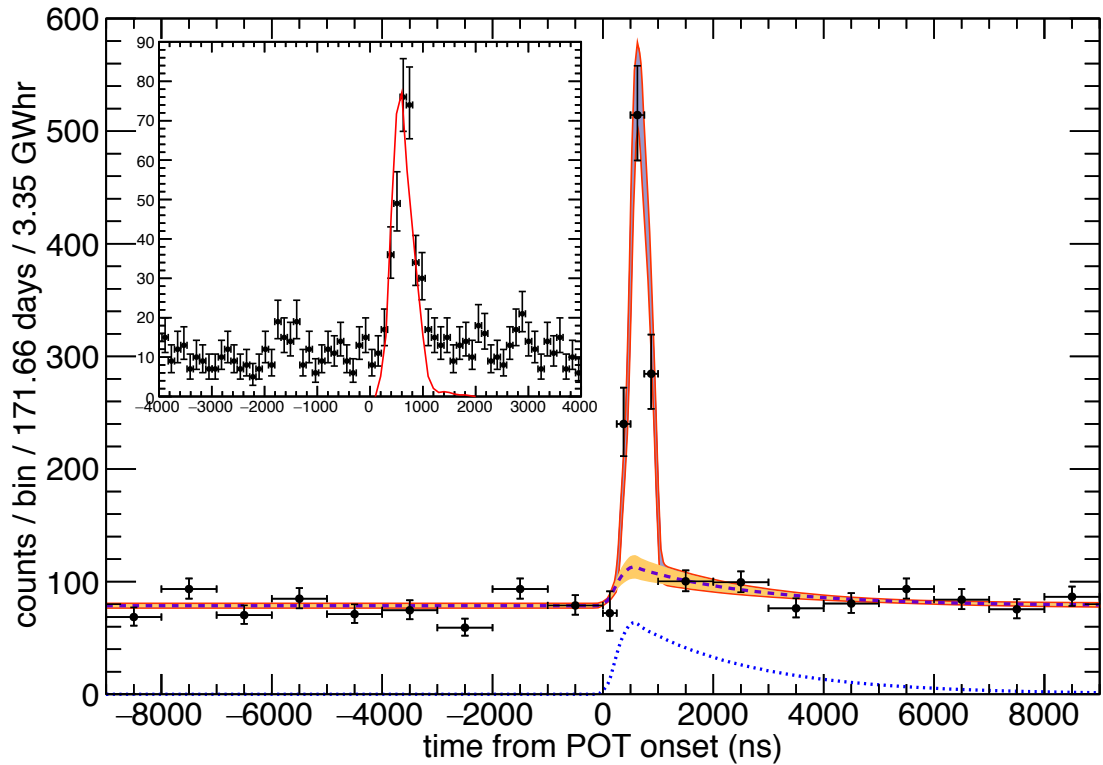

**Fig. S3**. Three-component unbinned fit to the arrival time of neutron-like events in EJ-301 scintillator cells (see text). Red lines delimit the one-sigma contour of the best-fit model. A dashed line indicates the best fit to NIN and environmental background components, a yellow band their one-sigma uncertainty. The presence of a non-zero NIN component is favored at the 2.9-sigma confidence level. However, the magnitude of this background is found to be negligible for a CEvNS search. A dotted line represents the predicted NIN component using the production rate calculated in (*57,58*). *Inset:* zoom-in using 100 ns bins. The red line is a normalized probability distribution function predicted by Geant4 for the arrival time of prompt neutrons contributing to the available 30-300 keV ionization energy region (Fig. S4). The simulation includes the time-profile of POT, provided by the SNS, and subsequent neutron production, moderation, and time-of-flight through 19.3 m of intermediate moderating materials (see text). This PDF is used to represent the prompt neutron component in our fits. The best-fit to its position agrees within errors (±168 ns) with the Geant4 prediction shown.

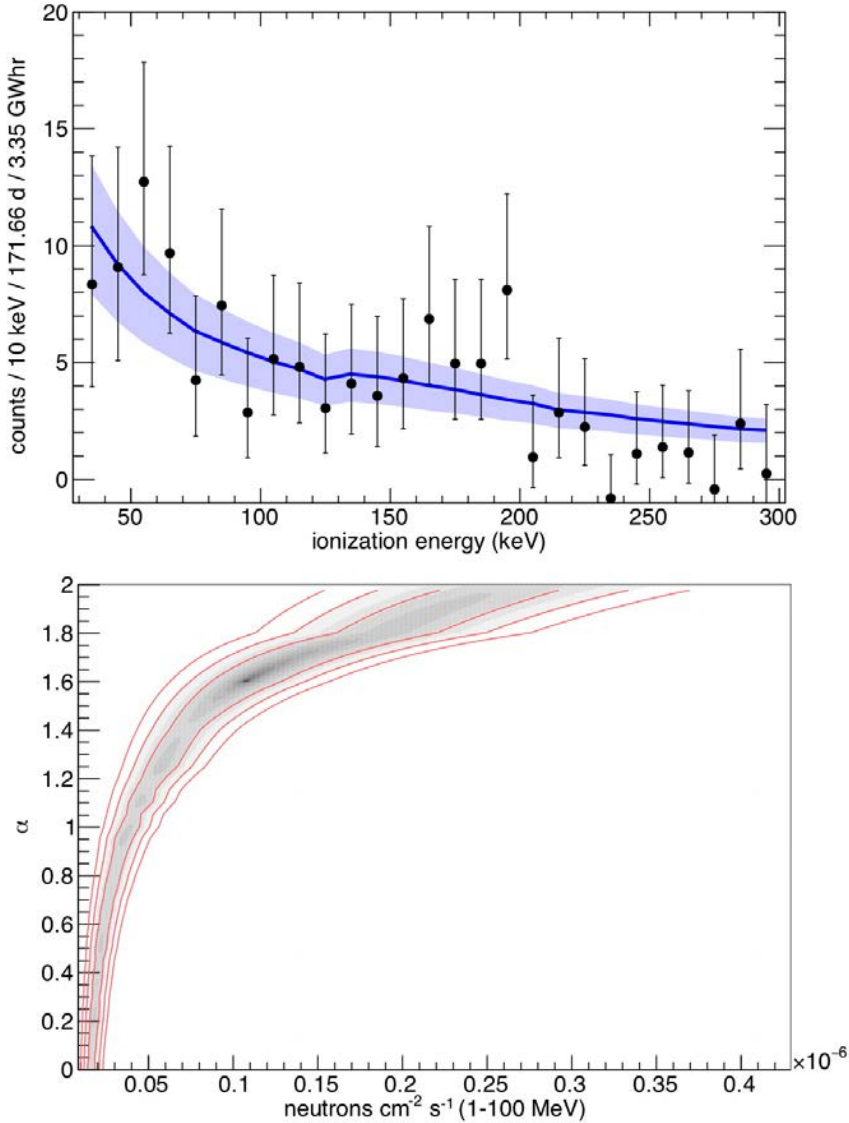

**Fig. S4**. *Top:* spectrum of energy depositions in EJ-301 from prompt neutrons (arrival times 200-1100 ns, Fig. S3), after environmental (random arrival) background subtraction. The blue line is the best-fit to the prompt neutron spectral models tested. A band encompasses all parameter combinations within the one-sigma region of the bottom panel. The deviation from an exponential originates in the model of response to low-energy proton recoils adopted (*55*). *Bottom:* fit quality for prompt neutron models at the detectors location having a spectral hardness $x^{-\alpha}$, where x is neutron energy in MeV. Overall fluxes in the 1-100 MeV range are plotted along the horizontal. Neutrons below 1 MeV have negligible probability of transporting across moderator layers in detector shielding, while contributing to the available EJ-301 energy range. Neutrons above 100 MeV are observed to have negligible flux in Scibath and Sandia Camera data, as well as Geant4 simulations. The best-fit parameters are $\alpha = 1.6$, and a flux of $1.09 \times 10^{-7}$ n / cm$^2$ s. This flux is compatible with Sandia Camera measurements in a nearby location (see text). Red contours delimit the 1-3 sigma confidence regions of the fit.

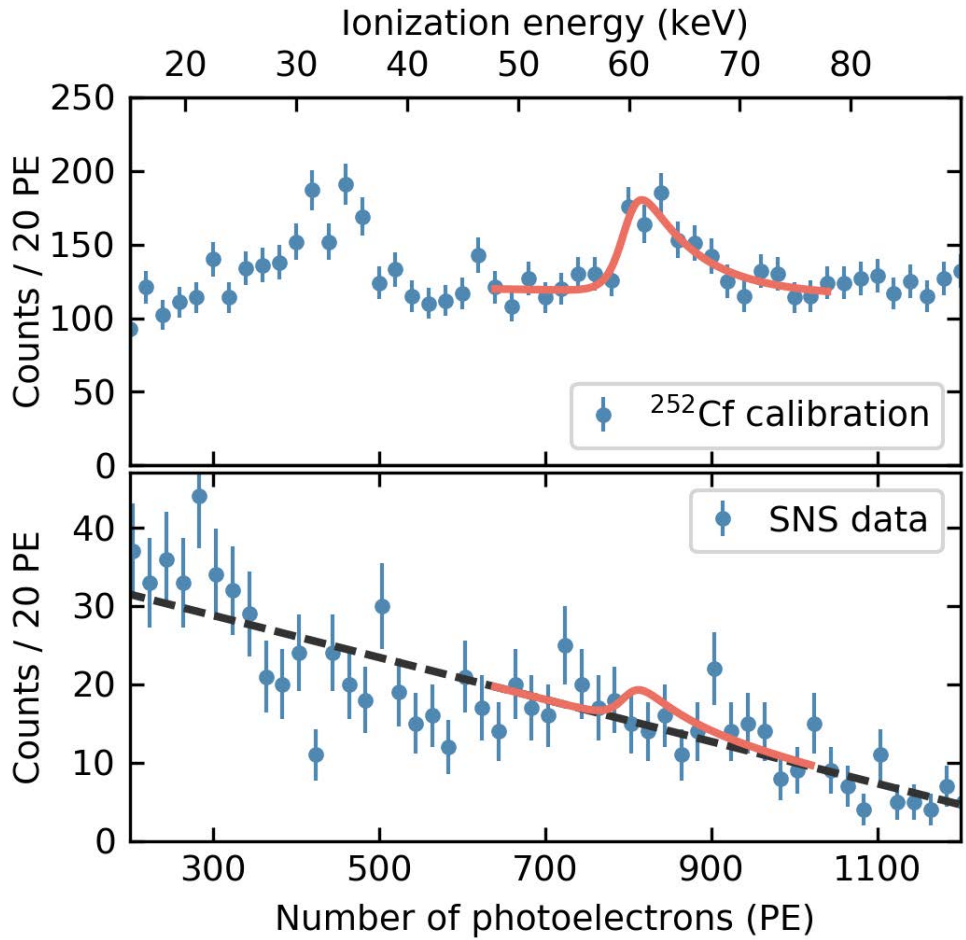

**Fig. S5**. *Top*: energy depositions in CsI[Na] during deployment of a $^{252}$Cf neutron source outside of the detector shielding, using self-triggering of the detector. A neutron inelastic scattering peak (57.6 keV) is visible at ~60 keV, with a second from the electron capture decay of $^{128}$I, at 31.8 keV. The small shift to a higher energy and "shark tooth" shape for the inelastic peak arise from the addition of recoiling nucleus and gamma de-excitation energies (*62*). This peak is correctly predicted in shape and rate by an MCNPX-PoliMi simulation (see text). The red line represents a fit to this region, using an *ad hoc* peak template and flat background. *Bottom:* Energy depositions in CsI[Na] during the 200-1100 ns arrival interval associated with prompt neutron arrival (Fig. S3), for all Beam ON periods collected (308.1 live-days). The red line shows the 90% C.L. maximum number of counts allowed by a fit using the same peak template and fitting window, and a simple background model (dashed black line). The best-fit number of counts under this peak is 3.9 ± 11.1, compatible with zero. An additional bound on the magnitude of the prompt neutron background can be extracted from the absence of an obvious inelastic peak (see text).

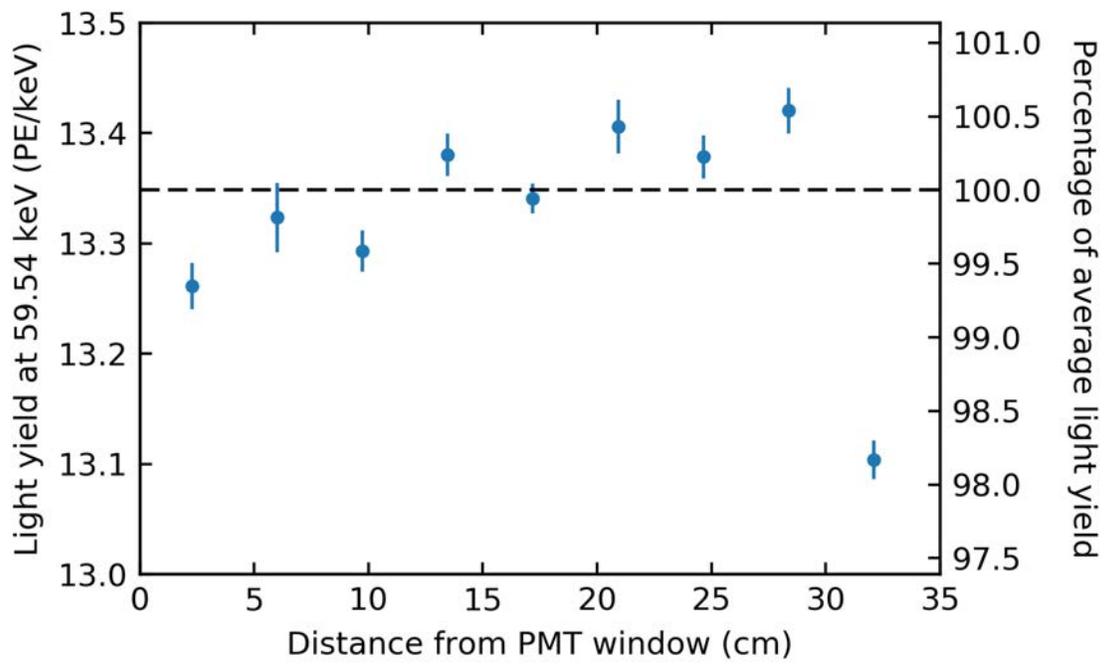

**Fig. S6**. Measurements of electron recoil light yield uniformity along the length of the CsI[Na] CEvNS detector (see text).

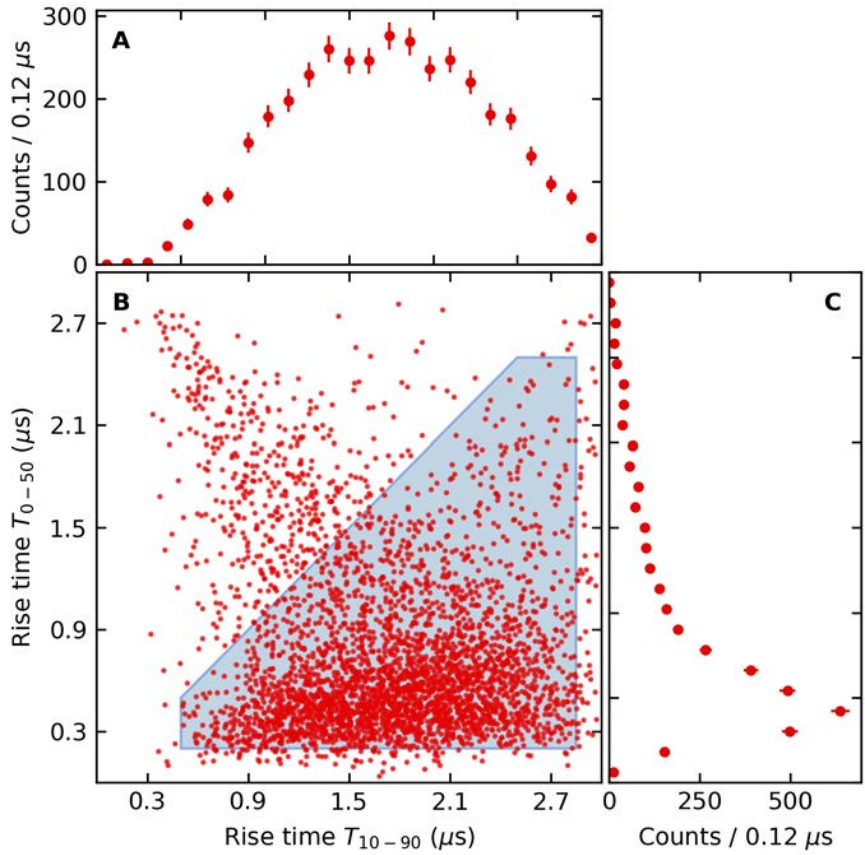

**Fig. S7**. Application of the integrated rise-time method (*50,51*) to the $^{133}$Ba data library of low-energy electron recoils, for events containing less than 50 PE. The choice of "Cherenkov" cut applied to these data is ≥ 8, a value optimized for the CEvNS search (see "Data Analysis"). Side panels (A,C) are data point projections on the corresponding axes, prior to cuts. Events in the blue-shaded region are accepted by the "Risetime" cut.

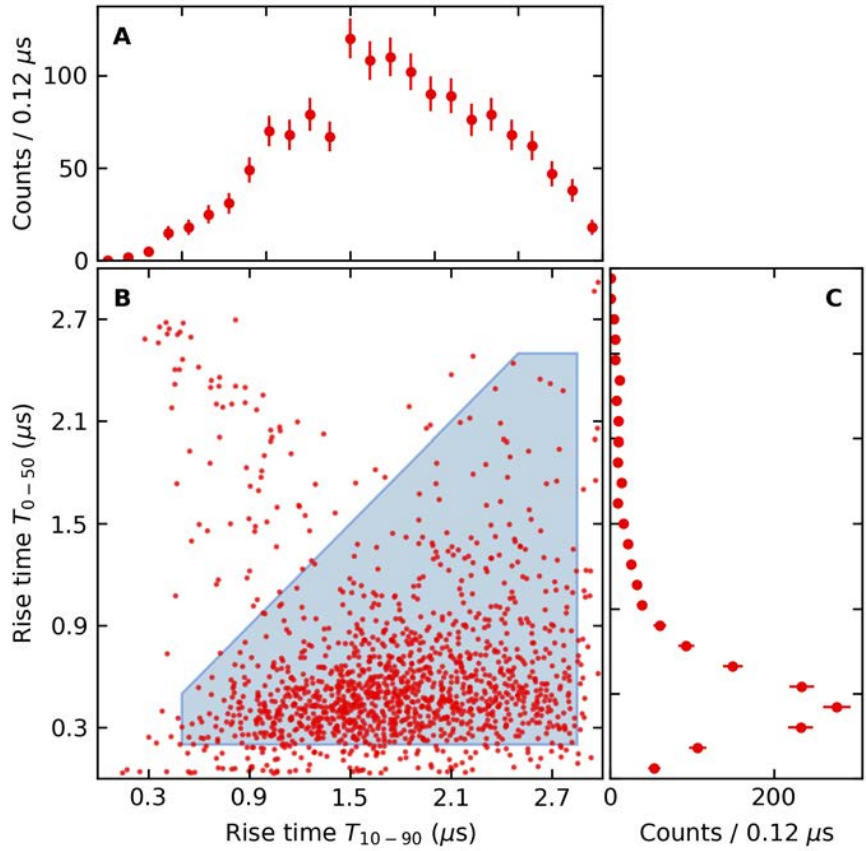

**Fig. S8.** Same as Fig. S7, for coincident (C) events registered during all Beam ON periods of the CEvNS search. Some of the slight differences between these rise-time distributions and those in Fig. S7 originate from variations in the fraction of events with misidentified onsets, a subset of them appearing above the diagonal in panel B. However, CsI[Na] scintillation decay constants are slightly different for low-energy nuclear recoils and electron recoils (*31, 67*): with the addition of more exposure, this property may provide a statistical discrimination of the nuclear and electron recoil components in the data.

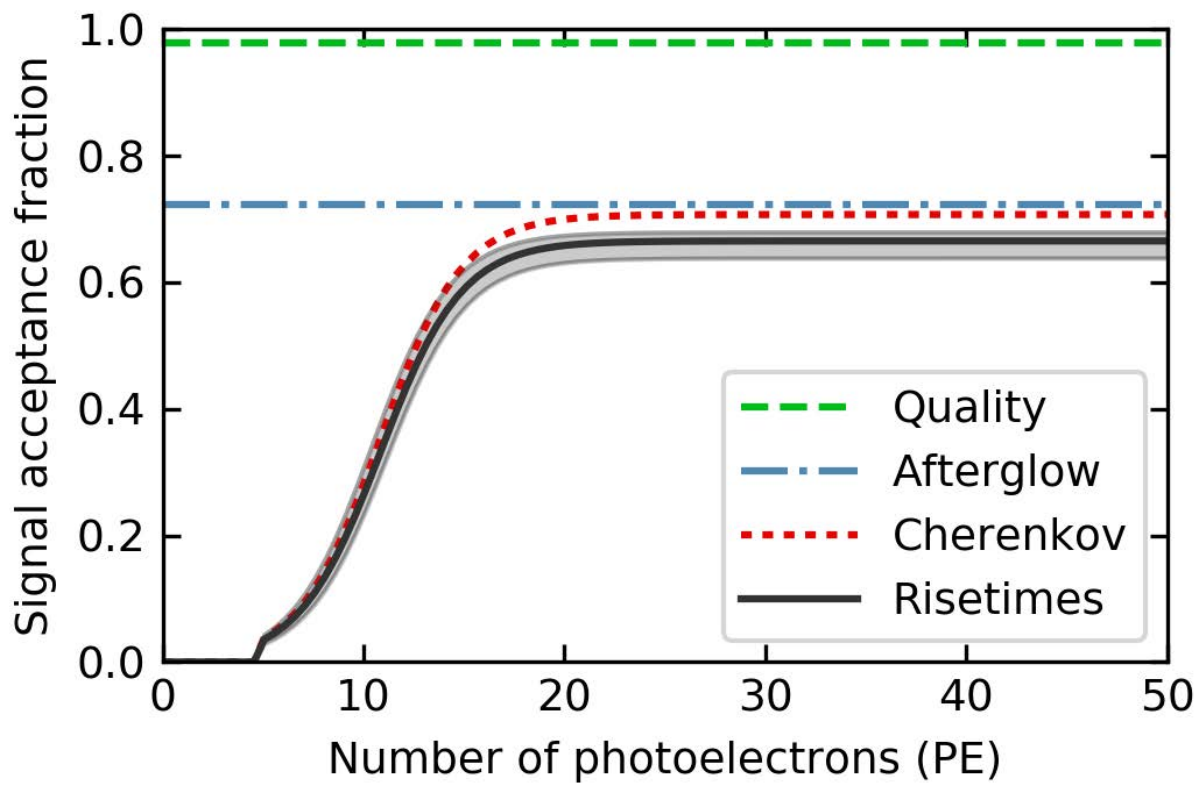

**Fig. S9.** Surviving fraction of CEvNS search data (Figs. 3 and S12), following cut choice optimization for a best signal-to-environmental background ratio (see "Data Analysis"). Cherenkov (≥ 8 peaks accepted) and Risetime (Figs. S7 and S8) cuts are defined using the $^{133}$Ba library. Afterglow (≤ 3 peaks accepted) and Quality cuts are defined using exclusively Beam OFF CEvNS search data (see "Data Analysis"). The uncertainty in this signal acceptance is expressed by a grayed band, and is dominated by the available $^{133}$Ba statistics. Using the electron light yield in Fig. S6 and best-fit quenching factor in Fig. S10, the onset of signal acceptance at 5 PE corresponds to a central value of nuclear recoil energy of 4.25 keV. The detectable fraction of total CEvNS signals as a function of CsI[Na] recoil energy threshold is given in (*31*).

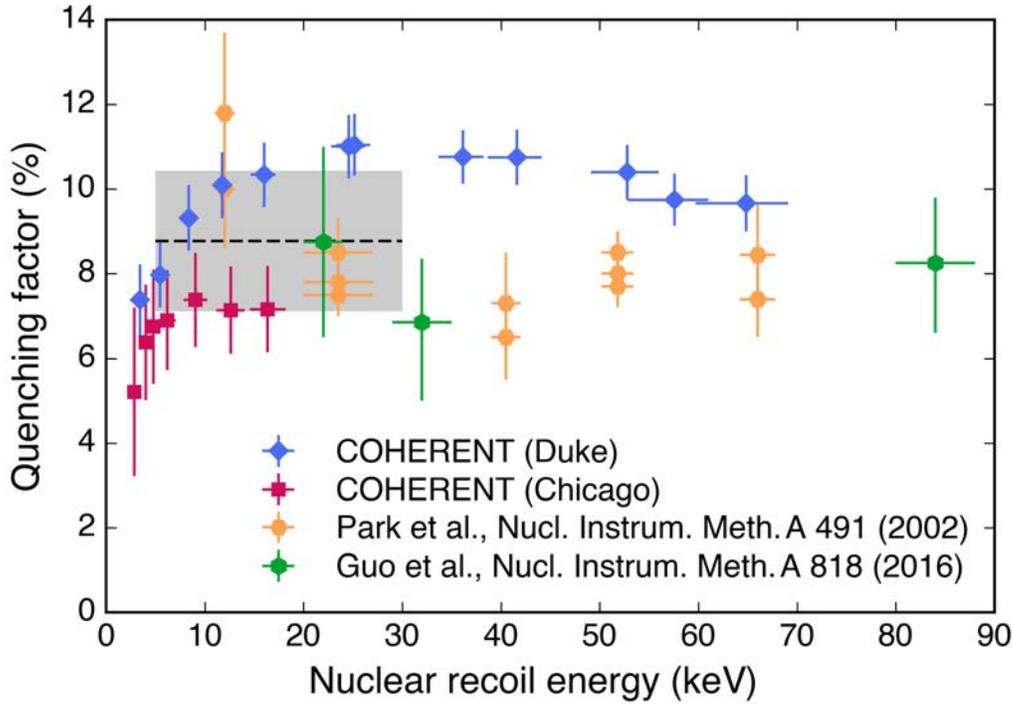

**Fig. S10.** Previous measurements of CsI[Na] quenching factor (*67, 68*), together with two new measurements performed within the COHERENT collaboration. These shared beam and target crystal, but differed in backing detectors, data acquisition, and approach to analysis. The grayed region spans the energy ROI for the present CEvNS search (~5-30 PE, Figs. 3 and S11). The reliability of semi-empirical QF models in this region being in question (*69*), we adopt the pragmatic approach of fitting all measurements in the ROI with a constant, weighting the experimental uncertainties shown in the plot (8.78 %, dashed line). Its uncertainty (± 1.66 %, vertical grayed range) is conservatively derived from the unweighted standard deviation of all data points included in the fit. We find no evidence in our data for an enhanced nuclear/electron recoil discrimination in CsI[Na], as claimed in (*68*).

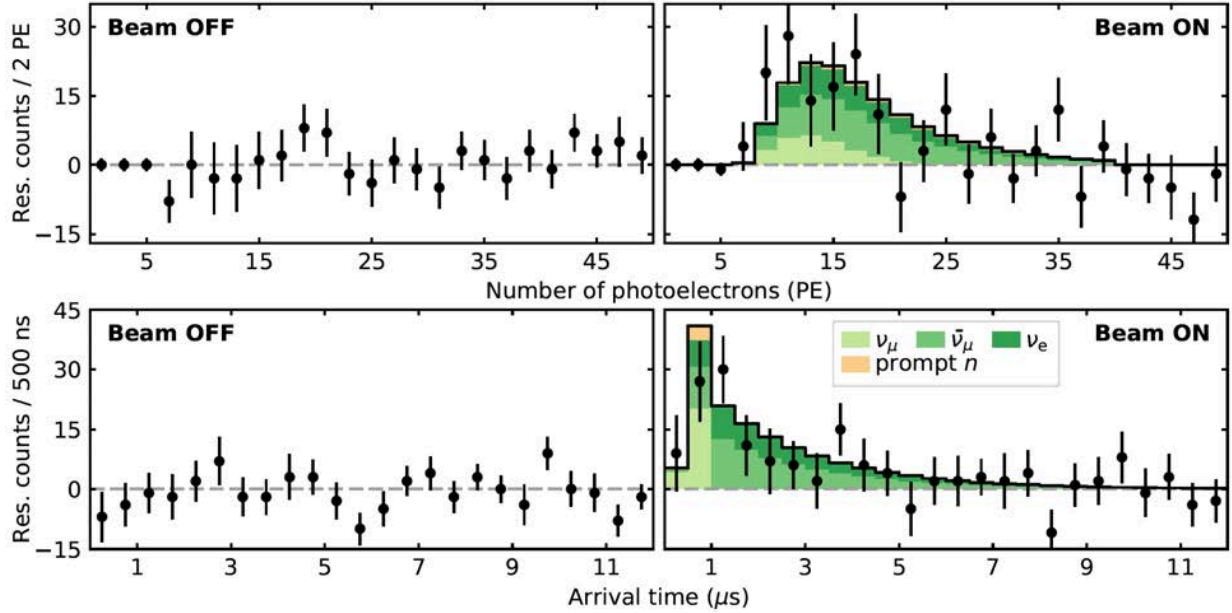

**Fig. S11.** Equivalent to Fig. 3 in the main text of this Report, from a parallel analysis pipeline (see text). Optimized choices of Cherenkov and Afterglow cuts for this analysis are ≥ 8 peaks accepted, and ≤ 4 peaks in pretrace accepted, respectively. Projections on energy (number of PE) are restricted to arrival times in the range 0-5 μs, and projections on time to PE ≤ 20. The CEvNS and prompt neutron predictions shown include the signal acceptance curve specific to this alternative analysis. The same good agreement with Beam ON residuals is observed, as well as an absence of CEvNS-like excess in Beam OFF data.

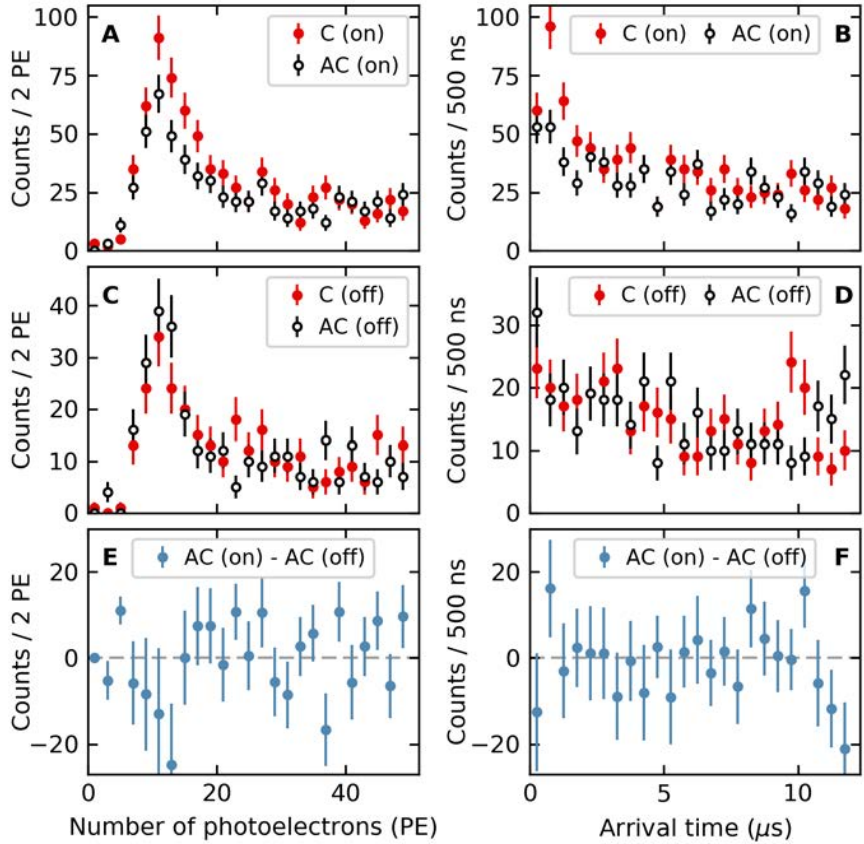

**Fig. S12.** *Panels A-D:* data passing all cuts in the first analysis pipeline, projected on energy (left column) for arrival times 0-6 μs, and on arrival time (right column) for $0 < PE \leq 30$. Beam OFF exposure was 153.5 live-days, Beam ON 308.1 live-days (data are not normalized to exposure). Error bars are statistical. Residuals in Fig. 3 are generated from the C-AC differences in the corresponding panels in this figure. The trend in time projections towards monotonically decreasing rates with increasing arrival time originates in events with a small afterglow component, able to pass all cuts. *Panels E-F:* exposure-corrected residuals between Beam ON and Beam OFF periods, for AC data (i.e., containing only steady-state environmental backgrounds). No significant deviation from zero is observed in either projection, demonstrating that modest changes in environmental gamma background in the "neutrino alley" from SNS operation are efficiently shielded by the >15 cm of lead around the CsI[Na] detector. These variations have a known origin in a nearby pipe carrying a steady flow of radioactive gas exhausts from the SNS Mercury Off-Gas Treatment System (MOTS, *78*). The emission is dominantly composed by 511 keV gammas (annihilation radiation), able to minimally affect the external muon veto panels (Fig. S1) facing the direction to the pipe, located a few meters away.

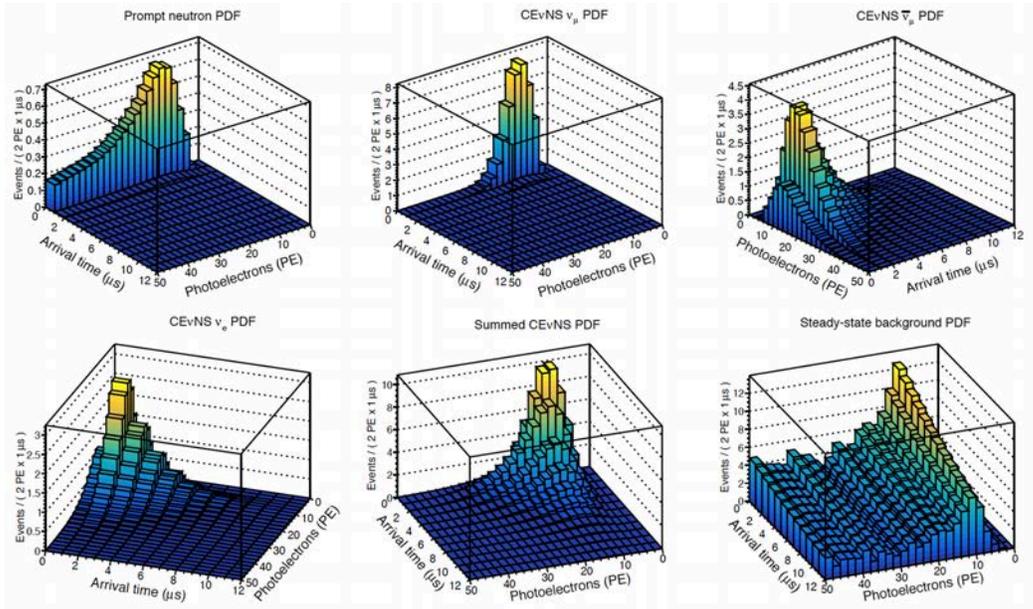

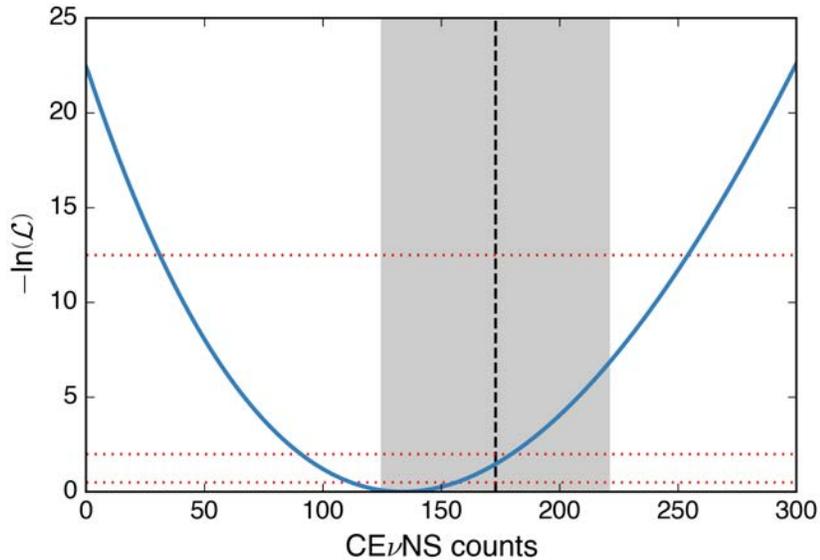

**Fig. S13.** *Top panels:* PDFs used in the 2-D (energy, time) fit described in the text. *Bottom:* Negative profile log-likelihood for the number of CEvNS events present in CsI[Na] data, using the model described in the text. Likelihood values are shifted so that the best-fit value from the data, 134 ± 22 CEvNS events, is drawn at 0. This result is within the 68% confidence band of the Standard Model prediction of 173 events, shown as a shaded region and a vertical dashed line. The 68%, 95%, and 99.9999% confidence levels (1 sigma, 2 sigma, and 5 sigma) of the fit are shown as ascending horizontal dotted lines. Comparison of log-likelihood values at counts of 0 and 136 indicates that the null hypothesis, corresponding to an absence of CEvNS events, is rejected at a level of 6.7-sigma, relative to the best fit.

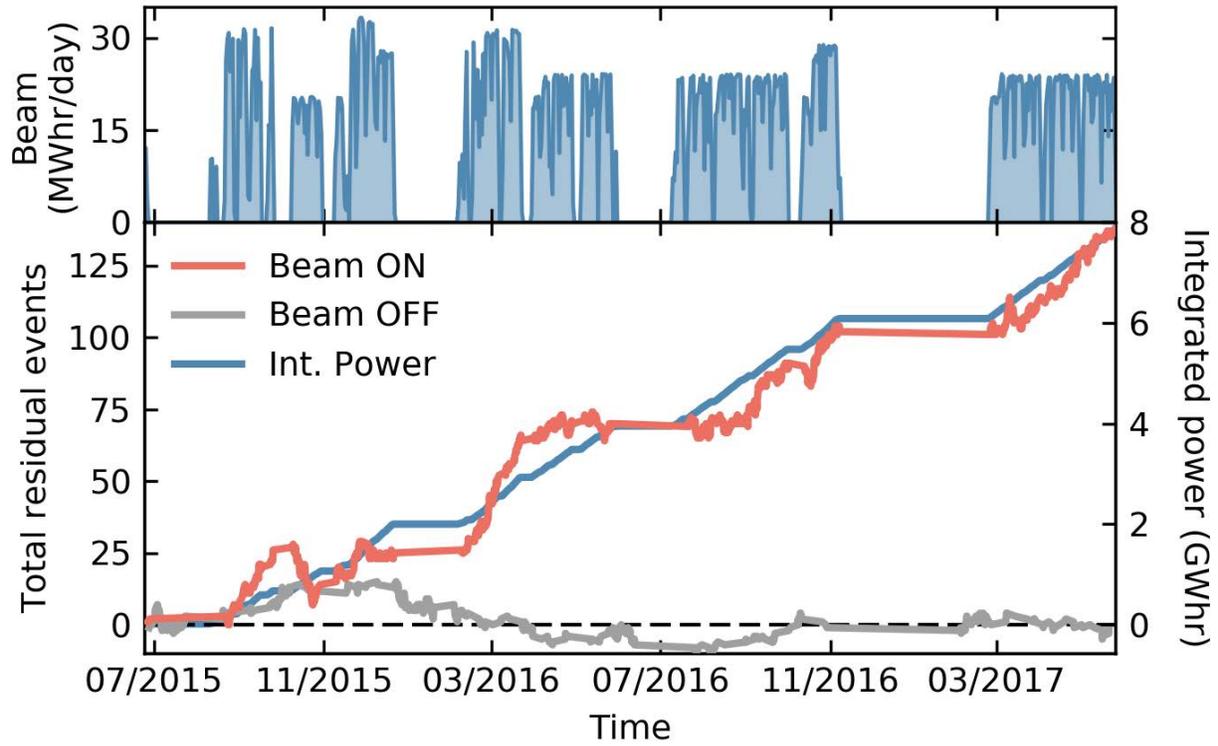

**Fig. S14.** *Top:* daily POT energy delivered by the SNS during data-taking. Long, planned SNS outages are visible. *Bottom:* Time evolution of the integrated number of counts in C-AC residual spectra (Fig. 3), for events with 0 < PE < 30 photoelectrons and arrival time of 0-6 μs (i.e., in the region where CEvNS is expected to dominate). The integrated rate for Beam ON periods (red) grows steadily with beam exposure (blue). Its behavior is consistent with a process entirely due to beam-induced events. No net change is observed for Beam OFF periods (gray). The integrated Beam OFF residual changes continuously due to frequent (daily) short unplanned outages, not all visible in the top panel. A Kolmogorov-Smirnov test finds no anomalous time-variation in the formation of the CEvNS-like excess in the C-AC beam ON residual.

**References (*36*)-(*84*)**